%% file: book.tex
\RequirePackage{etoolbox}
\csdef{input@path}%
{%
 {sty/},
 {img/},
}
\documentclass{elsevierbook}

\chardef\us=`\_

\usepackage{natbib}
\usepackage[pdftex,colorlinks,linkcolor=NavyBlue,citecolor=red]{hyperref}
\usepackage{totcount}

\newtotcounter{citnum} 
\def\oldbibitem{} \let\oldbibitem=\bibitem
\def\bibitem{\stepcounter{citnum}\oldbibitem}

\begin{document}

\Frontmatter

\Mainmatter
  \include{Chap4}

  \appendix

\Backmatter
%
%
%
%
%
%
%
%
%
 \bibliographystyle{elsarticle-harv}
%
%
%
%
%
%
%
%
%
%
%
%
%
%
%
%
%
\end{document}

%% file: Chap4.tex



\allowdisplaybreaks[1]

\newcommand{\fract}[2]{\leavevmode\kern.1em
          \raise.5ex\hbox{\the\scriptfont0 #1}\kern-.1em
    \raise.15ex\hbox{\the\scriptfont0 /}\kern-.08em\lower.25ex\hbox{\the\scriptfont0 #2}}

\newcommand{\ie}{\emph{i.e.}}
\newcommand{\eg}{\emph{e.g.}}
\newcommand{\cf}{{\emph{cf.}}}
\newcommand{\etc}{\emph{etc.}}

\newcommand{\cT}{c_{\scriptscriptstyle T}}
\newcommand{\zT}{z_{\scriptscriptstyle T}}
\newcommand{\cTe}{c_{\scriptscriptstyle Te}}
\newcommand{\CT}{C_{\scriptscriptstyle T}}
\newcommand{\CTe}{C_{\scriptscriptstyle Te}}
\newcommand{\kps}{$\rm km\,s^{-1}$}

\newcommand{\half}{{\textstyle\frac{1}{2}}}
\newcommand{\threeontwo}{{\textstyle\frac{3}{2}}}
\newcommand{\quart}{{\textstyle\frac{1}{4}}}
\newcommand{\re}{\mathop{\rm Re}\nolimits}
\newcommand{\im}{\mathop{\rm Im}\nolimits}
\newcommand{\sgn}{\mathop{\rm sgn}\nolimits}
\newcommand{\Ai}{\mathop{\rm Ai}\nolimits}
\newcommand{\Bi}{\mathop{\rm Bi}\nolimits}

\newcommand{\rd}{\mathrm{d}}
\newcommand{\rme}{e}
\newcommand{\rD}{\mathrm{D}}

\newcommand{\pderiv}[2]{\frac{\partial#1}{\partial#2}}
\newcommand{\pderivd}[2]{\frac{\partial^2#1}{\partial#2^2}}
\newcommand{\pderivt}[2]{\frac{\partial^3#1}{\partial#2^3}}
\newcommand{\pderivq}[2]{\frac{\partial^4#1}{\partial#2^4}}
\newcommand{\deriv}[2]{\frac{\rd#1}{\rd#2}}
\newcommand{\derivd}[2]{\frac{\rd^2#1}{\rd#2^2}}
\newcommand{\Deriv}[2]{\frac{\rD#1}{\rD#2}}
\newcommand{\Derivd}[2]{\frac{\rD^2#1}{\rD#2^2}}
\newcommand{\mpderiv}[3]{\frac{\partial^2#1}{\partial#2\partial#3}}

\newcommand{\kB}{k_{\scriptscriptstyle\rm B}}
\newcommand{\mH}{m_{\scriptscriptstyle\rm H}}

\renewcommand{\a}{{\mathbf{a}}}
\renewcommand{\b}{{\mathbf{b}}}
\newcommand{\bc}{{\mathbf{c}}}
\newcommand{\e}{\hat{\mathbf{e}}}
\newcommand{\f}{\mathbf{f}}
\newcommand{\g}{\mathbf{g}}
\newcommand{\h}{\mathbf{h}}
\renewcommand{\k}{\mathbf{k}}
\newcommand{\K}{\mathbf{K}}
\newcommand{\kk}{\mbox{\boldmath$\kappa$}}
\newcommand{\boldj}{\mathbf{j}}
\newcommand{\jp}{\mathbf{j}_\text{p}}
\newcommand{\bJ}{\mathbf{J}}
\newcommand{\m}{\mathbf{m}}
\newcommand{\n}{\mathbf{n}}
\newcommand{\boldd}{{\mathbf{d}}}
\newcommand{\boldu}{{\mathbf{u}}}
\newcommand{\boldv}{{\mathbf{v}}}
\newcommand{\boldw}{{\mathbf{w}}}
\newcommand{\boldl}{{\mathbf{l}}}

\newcommand{\boldr}{{\mathbf{r}}}

\newcommand{\x}{\mathbf{x}}
\newcommand{\A}{{\mathbf{A}}}
\newcommand{\B}{{\mathbf{B}}}
\newcommand{\C}{{\mathbf{C}}}
\newcommand{\E}{{\mathbf{E}}}
\newcommand{\F}{{\mathbf{F}}}
\newcommand{\G}{{\mathbf{G}}}
\newcommand{\Fh}{{}_2F_3}
\newcommand{\Fz}{{}_pF_q}
\newcommand{\I}{{\mathbf{I}}}
\newcommand{\J}{{\mathbf{J}}}
\newcommand{\N}{{\mathbf{N}}}
\newcommand{\X}{\mathbf{X}}
\newcommand{\M}{{\mathbf{M}}}
\newcommand{\Q}{{\mathbf{Q}}}
\newcommand{\R}{{\mathbf{R}}}

\newcommand{\T}{{\mathbf{T}}}
\newcommand{\U}{\mathbf{U}}
\newcommand{\V}{\mathbf{V}}
\newcommand{\W}{\mathbf{W}}
\newcommand{\Alpha}{\mathrm{A}}

\newcommand{\etaa}{\eta_\text{a}}
\newcommand{\etaH}{\eta_\text{\thth H}}
\newcommand{\taua}{\tau_\text{a}}
\newcommand{\tauH}{\tau_\text{\thth H}}
\newcommand{\Da}{\mathcal{D}_\text{a}}
\renewcommand{\DH}{\mathcal{D}_\text{H}}
\newcommand{\Ba}{\mathcal{B}_\text{a}}
\newcommand{\BH}{\mathcal{B}_\text{H}}
\newcommand{\ba}{\b_\text{a}}
\newcommand{\bH}{\b_\text{H}}
\newcommand{\pa}{p_\text{a}}
\newcommand{\pH}{p_\text{H}}

\newcommand{\tD}{\mbox{\boldmath$\mathsf{D}$}}
\newcommand{\tP}{\mbox{\boldmath$\mathsf{P}$}}
\newcommand{\tI}{\mbox{\boldmath$\mathsf{I}$}}



\newcommand{\Rm}{{\mathcal R}_m}

\newcommand{\vdot}{{\boldsymbol{\cdot}}}
\newcommand{\vcross}{{\boldsymbol{\times}}}
\newcommand{\grad}{\mbox{\boldmath$\nabla$}}
\newcommand{\bxi}{\mbox{\boldmath$\xi$}}
\newcommand{\bXi}{{\mbox{\boldmath$\Xi$}}}
\newcommand{\bpi}{\mbox{\boldmath$\pi$}}
\newcommand{\bmu}{\mbox{\boldmath$\mu$}}
\newcommand{\boldeta}{\mbox{\boldmath$\eta$}}
\newcommand{\bkappa}{\mbox{\boldmath$\kappa$}}
\newcommand{\bepsilon}{\mbox{\boldmath$\epsilon$}}
\newcommand{\bdelta}{\mbox{\boldmath$\delta$}}
\newcommand{\bpsi}{\mbox{\boldmath$\psi$}}
\newcommand{\bDelta}{{\mathbf{\Delta}}}
\newcommand{\bPhi}{{\mathbf{\Phi}}}
\newcommand{\bOm}{{\mathbg{\Omega}}}
\newcommand{\diag}{\mathop{\rm diag}}
\newcommand{\thth}{\hspace{1.5pt}}
\newcommand{\curl}{\grad\vcross}
\newcommand{\Curl}{\grad\vcross\thth}
\renewcommand\div{\grad\vdot}
\newcommand\Div{\grad\vdot\thth}
\newcommand{\gradperp}{\grad_{\!\!\perp}}
\newcommand{\nablaperp}{\nabla_{\!\!\perp}}
\newcommand{\gradp}{\grad_{\!\text{p}}}
\newcommand{\nablap}{\nabla_{\!\text{p}}}

\newcommand{\kperp}{k_{\scriptscriptstyle\!\perp}}
\newcommand{\vperp}{v_{\scriptscriptstyle\!\perp}}
\newcommand{\vpar}{v_{\scriptscriptstyle\parallel}}
\newcommand{\uperp}{u_{\scriptscriptstyle\!\perp}}
\newcommand{\upar}{u_{\scriptscriptstyle\parallel}}
\newcommand{\kperpz}{k_{\scriptscriptstyle\!\perp0}}
\newcommand{\kpar}{k_{\scriptscriptstyle\parallel}}

\newcommand{\kp}{k_{\scriptscriptstyle\parallel}}
\newcommand{\kpzero}{\kappa_0}

\newcommand{\ct}{c_{\scriptscriptstyle T}}
\newcommand{\tB}{\theta_{\scriptscriptstyle B}}

\newcommand{\bv}{Brunt-V\"ais\"al\"a}

\newcommand{\ri}{{i}}

\newcommand{\ts}{\textstyle}
\newcommand{\scsc}{\scriptscriptstyle}

\newcommand{\calA}{{\mathcal{A}}}
\newcommand{\calB}{{\boldsymbol{\mathcal{B}}}}
\newcommand{\calL}{{\mathcal{L}}}
\newcommand{\calD}{{\mathcal{D}}}
\newcommand{\calU}{{\mathcal{U}}}

\newcommand{\kpz}{\kappa_z}

\newcommand{\eperp}{\e_{\scriptscriptstyle\!\perp}}
\newcommand{\epar}{\e_{\scriptscriptstyle\parallel}}

\renewcommand{\leq}{\leqslant}  \renewcommand{\le}{\leqslant}
\renewcommand{\geq}{\geqslant}  \renewcommand{\ge}{\geqslant}


\cleardoublepage
\begin{frontmatter}
\setcounter{chapter}{3}
\chapter{MHD Waves in Homogeneous and Continuously Stratified Atmospheres }\label{chap4}

\begin{aug}
\author[addressrefs={adPC}]%
  {\fnm{Paul S.}   \snm{Cally}}%
\author[addressrefs={adTB}]%
  {\fnm{Thomas J.}   \snm{Bogdan}
  }%
\address[id=adPC]%
  {School of Mathematics, Monash University, Clayton, Victoria 3800, Australia}%
\address[id=adTB]%
  {National Solar Observatory, 3665 Innovation Drive, Boulder CO 80303, USA}
\end{aug}

\begin{abstract}
 The basic equations, concepts, and modes of linear, ideal, MHD waves -- slow, Alfv\'en and fast -- are set out and generalised to gravitationally-stratified atmospheres. Particular attention is devoted to mode conversion, wherein  
 the local behavior of a global wave changes from one mode to another in passing through particular atmospheric layers. Exact solutions are explored where available. Eikonal methods -- WKBJ and ray theory -- are described. Although our emphasis is on the theoretical underpinning of the subject, the solar atmospheric heating implications of fast/slow and fast/Alfv\'en conversions are discussed in detail.
\end{abstract}

\begin{keywords}
\kwd{MHD}
\kwd{Waves}
\kwd{Sun: atmosphere}
\end{keywords}

\end{frontmatter}

\section{Introduction} \label{sec:intro}

This chapter discusses the nature of waves in magnetized
astrophysical fluids. Owing to its proximity and visibility,
we are primarily concerned with the optically-thin
solar atmosphere.
However, our results and findings are widely applicable to astrophysical
settings where the plasma admits a continuum or fluid-dynamical description.

The subject of magneto\-hydro\-dynamic (MHD) waves is not without its intrinsic
subtleties. They result from the linearization
of the fully nonlinear magneto\-fluid dynamics about an 
appropriate equilibrium state. 
In this sense, they obtain formally 
in the limit of infinitesimal amplitudes, i.e., when quadratic terms may be neglected
compared to the leading-order linear terms. On the other hand, every observation is
predicated upon the presence of a finite-amplitude disturbance in an astrophysical fluid.
In this fashion, MHD waves are mathematical idealizations, which 
hopefully capture the essential
behavior of small-amplitude fluctuations in actual magnetized 
astrophysical plasmas. The fundamental
issue is which fluctuations are of sufficiently small amplitude, and which are not. 
Stated differently: under what conditions may the second-order quadratic terms be safely
discarded?

This question must be addressed on a case-by-case basis. In
turn it poses a deeper and more perplexing question: what aspects or components
of a given dynamical MHD flow are to be identified with MHD waves? And to what then
do we ascribe the remainder of the time-dependent flow? This is a topic of active
investigation and we shall be content to simply offer a few useful insights in what follows. 

The applicability of the MHD fluid description of waves is bounded in both
temporal frequency and spatial wavelength, or wavenumber. At high frequencies and small
wavelengths the particulate nature of the plasma must be taken into account. The
plasma frequency and gyro-frequencies of the electrons and ions serve to mark this
boundary between particle and collective fluid behaviors. Even collisionless
plasmas may be treated as a fluid, provided care is taken in developing an accurate equation
of state. Dissipative transport processes may be accommodated within the fluid picture
when frequencies and wavelengths are comparable to 
collective relaxation times and lengths (Appendix \ref{app:thermo}).
At the other extreme of low frequencies and large wavelengths, one must
take care in identifying the appropriate equilibrium state. In some instances it is 
necessary to allow this equilibrium to evolve slowly in time or to be treated as a random
medium. Such extensions of the fluid MHD wave description are beyond the scope of this 
chapter.

MHD waves may be modified and duly influenced by physical processes like rotation, 
buoyancy, and steady fluid motions, for example. The resulting menagerie of hybrid waves
has spawned a vast and sometimes bewildering nomenclature. In a stellar atmosphere, the
radiation field frequently exerts the dominant influence on MHD waves. It not 
only provides a dissipation mechanism, but in hot stars it may also modify the propagation
characteristics of the waves (Appendix \ref{app:RMHD}).

\subsection{Magneto-acoustic-gravity (MAG) Waves in the Combined Photosphere/Chromosphere}

A rich variety of oscillations and waves are observed between the base 
of the solar corona and the surface of the Sun \citep{00bogdan,Rut03aa,06bogdan,15khomenko,
16lohner,21srivastava,JesJafKey23aa}.
The solar 
convection and its overshoot are the primary mechanical sources that generate
these disturbances. Intermittent flares of all scales throughout,
and above, this roughly 2 Mm-thick optically-thin atmospheric layer
are also wave sources. Generally speaking, these waves are able to partially pass through
the upper and lower boundaries of the combined photosphere and chromosphere. 
In both directions they encounter a rapid increase in the average plasma temperature and
characteristic vertical (radial) density scale-height. 

Indeed, what sets the 
combined photosphere/chromosphere apart from the neighboring corona and 
convection zone is its low temperatures ($T \approx 4000-6000$ K)
and small scale-heights ($H_\rho \approx 150-200$ km). These attributes pose 
serious challenges to modeling magneto-acoustic-gravity (MAG) waves. Structures like granules, 
pores, sunspots and the magnetic network exhibit horizontal scales that are very much
larger than $H_\rho$. This effectively breaks the translational symmetry that might 
otherwise permit Fourier analysis in the vertical (i.e., radial) direction. As we shall
presently demonstrate, this in turn necessitates the solution of ordinary differential
equations (ODEs) with non-constant coefficients.

A second complication is provided by the convective overshoot. 
The photosphere/chromosphere may be regarded 
as a solar `surf-zone' where the inertia of episodic convective upflows carries
them far into the overlying stable atmosphere before buoyancy breaking 
eventually halts their
progress \citep{95carlsson,11uitenbroek,13criscuoli}.
Spicules and umbral flashes are 
familiar observational signatures of these processes.
Mass balance ensures a gradual subsidence of the dense material back to the
convection zone. Therefore, the photosphere/chromosphere is {\em dynamic} in a
chaotic sense. Flares and coronal transients are additional sources of chaotic
dynamics. The photospheres and chromospheres of stars with convective
envelopes must sustain similar conditions. Finally, as Lighthill, Proudman,
Parker and Kulsrud have all pointed out, this chaotic dynamics will in turn generate waves and
oscillations in situ \citep{52lighthill,52proudman,53parker,54lightill,55kulsrud}. See \citet{GolMurKum94aa} and references therein for more recent developments.

Coherent waves and oscillations with sufficiently large wavelengths and long periods will
emerge from and propagate through these optically-thin stellar surf-zones. For the Sun, 
the dominant temporal periods are on the order of a few minutes. Horizontal wavelengths
cover a much broader range from fractions of a solar radius down to a few tenths of a
second of arc (comparable to the density scale-height in the photosphere). As Dewar noted,
large-scale coherent waves/oscillations sample the physical conditions in the 
chaotic photosphere/chromosphere surf-zone in a fashion that depends upon their 
individual nature and they adjust their properties accordingly \citep{70dewar,71dewar}. 
The `wave mean' of a physical
quantity, like the (vector) magnetic field, the adiabatic compressibility, or
advective flow, for example, may be distinct from simple
(i.e., unweighted) spatial and temporal averages.
This adds further subtlety to the analysis and complicates 
spectropolarimetric inversions.

We shall simply refer to periodic fluctuations detected in the solar 
photosphere/chromosphere as MAG waves or oscillations because the principal restoring forces are some combination of magnetic pressure/tension, plasma compressibility, and gravitational stratification/buoyancy. Except in special circumstances, radiative transfer, fluid viscosity, thermal conduction, ohmic dissipation, and solar rotation will
have a lesser, or secondary, influence. 

The first detection of solar MAG waves came in the late 1960s \citep{69beckers,74schultz}. Since then, they have
received a rapidly expanding amount of attention and careful study. This has been 
facilitated by advances in instrumentation -- larger telescope apertures, faster CCDs,
precision spectropolarimetry, broad spectral access from space -- and computational 
advances. Indeed, entire disciplines, such as astero- and helioseismology emerged and
have now become mature areas of study. Offshoots, like the seismology of spots and 
coronae are flourishing.

For several reasons, attention has focused on waves and oscillations with
periods between 1 and 10 minutes. These waves exhibit characteristic ridge
or ring structures in traditional $k$--$\omega$ power-spectra diagrams. At both
larger and smaller oscillation periods these coherent structures fade into an
incoherent continuum of oscillations with random phases, amplitudes, and wavevectors. Apropos the question raised in the introduction, these power spectra contain large contributions from the turbulent convection in addition to waves. 

We can do no better than simply point the reader to several current reviews on
these topics \citep{02jcdc,16basu,21brown}. 
However, a few cautionary remarks will be helpful for what follows.
{\em First}, there are clear differences between the periodic fluctuations 
detected in strong (and perhaps laminar, unipolar) magnetic fields (i.e., the umbrae of spots), and the surrounding quiet Sun. No doubt, there are ample `salt-and-pepper' magnetic fields present throughout the quiet Sun. In contrast to the umbral magnetic fields, they are turbulent, bipolar, randomly-oriented, and possibly fibril (or better, intermittent) in nature. Inclined penumbrae lie somewhere between these two extremes. This observational dichotomy suggests, \`a la Dewar, that the waves/oscillations take rather \emph{different} averages of these two extreme states of solar surface magnetism. \emph{Second}, it is necessary
to distinguish between the bona fide propagation of a disturbance and the sequential
emergence of a disturbance's wave-front through a particular atmospheric surface
where a spectral diagnostic is formed. A good
case in point is the phenomenon of running penumbral waves. \emph{Third}, the theory of MAG waves invokes the mathematical limit of letting the wave amplitude tend to zero, whilst any observed fluctuation must necessarily have a \emph{finite} amplitude. Chromospheric umbral flashes, for example, are most certainly nonlinear wave-trains that
have steepened into shocks. 

\subsection{Overview}
In this chapter we first introduce the basic equations of MHD waves and then 
derive their simple iconic forms (slow, Alfv\'en and fast) in a uniform,
translationally-invariant, plasma. These iconic forms have the distinct pedagogical
advantage that the governing partial differential equations (PDEs) factor into three
distinct decoupled forms, or wave-modes.  

We then focus primarily on what is perhaps the dominant departure from this simple picture pertaining to the low solar atmosphere: gravitational stratification in the vertical direction, giving rise to magneto-acoustic-gravity (MAG) waves which may only {\em locally} adhere to the fast-Alfv\'en-slow trichotomy. Indeed, the essential point is that
the governing PDEs do not in general factor, or decouple, unless the equilibrium
possesses one or more translational symmetries. 

Instead, one may be able to identify spatial
domains where one of the wave-modes is only weakly-coupled to the remainder.
{\em Globally}, these MAG waves exhibit mixed characteristics at different locations.
This may be conveniently regarded as the result of wave mode-conversion processes. 
We shall presently enumerate and illustrate the various mode-conversion processes.

In general, however, when no spatial symmetries are present in the equilibrium,
the governing wave PDEs do not decouple and the characterization and classification of the
resulting MAG waves is distinctly equilibrium-dependent. 
Such problems are of course the most germane to astrophysical situations.
They are invariably treated by numerical methods. 

Other chapters in this volume address important deviations from this elementary scenario, such as flux tubes (Chapter 5), partial ionization (Chapter 6) and nonlinearities (Chapter 8). All of these aspects contribute to the important topics of coronal heating (Chapter 10) and solar wind acceleration (Chapter 11).

\section{MHD Equations}\label{sec:mhd eqns}

The single-fluid MHD equations are commonly expressed in terms of the density $\rho$, plasma velocity $\boldv$, magnetic induction $\B$, current density $\boldj=\mu^{-1}\Curl\B$ (current per unit area), and gravitational acceleration $\g$. In solar physics, $\B$ is normally called the magnetic field, though strictly, in a macroscopic medium, the magnetic field $\mathbf{H}$ is related to the magnetic induction $\B$ via a constitutive relation $\B=\mu\mathbf{H}$, but in astrophysical plasmas $\mu$ is accurately represented by its vacuum value $\mu_0=4\pi\times10^{-7}$ $\rm H\,m^{-1}$ in SI units, and $\mathbf{H}$ plays no independent role.

\subsection{Conventional Form}
Mass conservation is expressed by
\begin{equation}\label{cty}
\Deriv{\rho}{t}+\rho\Div{\bf v}=0,
\end{equation}
familiar from hydrodynamics, where
\begin{equation}
\Deriv{\phantom{t}}{t}=\pderiv{\phantom{t}}{t}+
{\bf v}\thth\vdot\grad \label{comov}
\end{equation}
is the \emph{comoving derivative.}

Similarly, the momentum equation also carries over from neutral fluid theory,
\begin{equation}
\rho \Deriv{\boldv}{t}=-\grad p+{\boldj\,\vcross\, \B}+\rho\,\mathbf{g}
+\eta_v\nabla^2{\bf v}+\left(\zeta+\frac{1}{3}\eta_v\right)\grad\left(
\Div{\bf v}\right),  \label{mmntm_unifvisc}
\end{equation}
though with the addition of the \emph{Lorentz force} ${\boldj\,\vcross\, \B}$, which may be decomposed into magnetic pressure gradient and magnetic tension components $-\grad(B^2/2\mu)+\Div(\B \B/\mu)$. The gravitational acceleration $\mathbf{g}=-\grad\phi$ is normally specified externally, but can be calculated self-consistently using the Poisson equation $\nabla^2\phi=4\pi G\rho$ for a self-gravitating system.
For completeness, the coefficients of dynamic and bulk viscosity $\eta_v$ and $\zeta$ have been included, though these are negligible in most solar contexts and will be dropped hereafter. The momentum equation describes how the magnetic field affects the plasma's motion.

Conversely, the flow of plasma also affects the magnetic field, as described by the induction equation
\begin{equation}
\pderiv{\B}{t}=\curl\left(\boldv\,\vcross\,\B\right)
-\curl\!\left(\eta\Curl \B\right),  \label{inductiona}
\end{equation}
where $\eta=1/\mu\sigma$ is the \emph{magnetic diffusivity} and $\sigma$ is the electrical conductivity. If $\eta$ is
uniform, this is more commonly expressed in the form
\begin{equation}
\pderiv{\B}{t}=\curl\!\left(\boldv\,\vcross \,\B\right)
+\eta\nabla^2{\B},     \label{inductionb}
\end{equation}
from which the diffusive nature of $\eta$ is apparent. We shall also neglect the magnetic diffusivity, restricting attention to so-called \emph{ideal MHD}.

\subsection{Conservation Form}
For numerical purposes, and also when deriving the jump conditions for shock waves (see Chapter 9), the MHD equations are more useful in conservation form $\partial\mbox{`density'}/\partial t+\mbox{div`flux'}=0$, where `density' is some scalar or vector quantity per unit volume and `flux' is a vector or dyadic generalized flux of the same quantity.

For example, the mass conservation equation for density $\rho$ takes the form
\begin{equation}
    \pderiv{\rho}{t}+\Div(\rho\boldv)=0,
\end{equation}
with $\rho\boldv$ being the mass flux per unit volume.
Similarly, momentum conservation may be expressed as
\begin{equation}
    \pderiv{\rho\boldv}{t}+\Div\mathbf{T}=-\rho\grad\phi,
\end{equation}
where $\phi$ is the gravitational potential such that $\g=-\grad\phi$, and the non-zero right hand side indicates that momentum is in fact not conserved in an external gravitational field. The dyadic 
\begin{equation}
    \mathbf{T} = \rho\boldv\boldv + \left(p+\frac{B^2}{2\mu}\right)\I-\frac{\B\B}{\mu}
\end{equation}
is the total stress tensor, where $\I$ is the $3\times3$ identity. The ideal induction equation may also be written in conservation form,
\begin{equation}
    \pderiv{\B}{t}+\div\left(\boldv\B-\B\boldv\right)=0,
\end{equation}
since magnetic flux is conserved. Finally, the total energy equation expresses the Eulerian time derivative of the local energy density $U=\half\rho v^2+p/(\gamma-1)+B^2/2\mu+\rho\phi$ (kinetic+thermal+magnetic+gravitational) in terms of the divergence of the total energy flux $\f$
\begin{equation}
    \pderiv{U}{t}+\Div\f= 0  
\end{equation}
where
\begin{equation}
\f = \left( \half\rho\thth v^2+\frac{\gamma}{\gamma-1}p+\rho\phi\right)\!\boldv - \frac{1}{\mu} (\boldv\vcross\B)\vcross\B~, \label{f}
\end{equation}
and $\gamma$ is the ratio of specific heats. This flux corresponds to the sum of the advected kinetic, thermal and gravitational energies, the rate of working of the pressure force $p\boldv$, and the electromagnetic (Poynting) flux, $\mu^{-1}\E\vcross\B$, where $\E=-\boldv\vcross\B$ (in ideal MHD) is the electric field.

Often the first law of thermodynamics is invoked to provide an alternate 
but equivalent form of these
last two equations couched in terms of conservation of specific entropy $\eta$ of the fluid,
\begin{equation}\label{DsDt}
    \rho\, T \,\Deriv{\eta}{t} = -\mathcal{L},
\end{equation}
where $\mathcal{L}$ is the net energy loss per unit volume (outgoing minus incoming), made up of radiation, conduction, viscous, Joule, etc.~terms (Appendices \ref{app:thermo}, \ref{app:RMHD}). In the absence of heating or loss, specific entropy and energy are conserved following the motion.

\section{MHD Equilibria}\label{sec:eqbm}
\subsection{Lorentz Force and Equilibrium}
Setting $\boldv=\mathbf{0}$ in the momentum equation (\ref{mmntm_unifvisc}) results in the equation of magneto\-hydro\-static (MHS) equilibrium,
\begin{equation}
\mathbf{0}=-\grad p+{\boldj\,\vcross\, \B}+\rho\,\mathbf{g}.  \label{eqbm}
\end{equation}
In all but some very simple symmetric scenarios this equation is challenging to solve. 
The essential difficulty is the Lorentz force must be the gradient of a scalar (pressure) 
in every two-dimensional manifold perpendicular to {\bf g}. This is difficult to arrange 
in principle \citep{85low,86bogdan,99neukirch,05low},
and as Parker has demonstrated, essentially impossible to achieve in practice 
\citep{94parker}. 

To appreciate this equation, one must understand the Lorentz force (actually a force per unit volume), $\F_\text{L}=\boldj\,\vcross\, \B=\mu^{-1}(\Curl\B)\vcross\,\B$. Taking a curl and then a cross product of a known magnetic field in one's head, especially if expressed in cartoon form, is probably beyond most of us, so how can we understand the Lorentz force for a given sketched $\B$?

Using standard vector identities, we may recast
\begin{equation}\label{FL}
    \F_\text{L} =  \B\,\vdot\grad\,\B/\mu - \grad(B^2/2\mu),
\end{equation}
which can be interpreted respectively as a \emph{magnetic tension} force along field lines, and a \emph{magnetic pressure} force $-\grad p_\text{mag}$ where $p_\text{mag}=B^2/2\mu$ is the magnetic pressure. Hence Equation (\ref{eqbm}) can be rewritten as 
\begin{equation}
\mathbf{0}=-\grad \left(p+p_\text{mag}\right)+\B\,\vdot\grad\,\B/\mu+\rho\,\mathbf{g}, \label{eqbmAlt}
\end{equation}
making the equilibrium balance between total pressure $p+p_\text{mag}$, tension and gravity more intuitive. We mention in passing that $B^2/2\mu$ is also the magnetic energy density,
and the Lorentz force can also be expressed as the divergence of the symmetric Maxwell
stress tensor.

In the solar corona, magnetic energy density typically exceeds thermal energy density by an order of magnitude, so it a reasonable approximation to set $\F_\text{L}=\mathbf{0}$ to obtain a (near) equilibrium. Any $\B$ satisfying this condition is called a \emph{force free field}, of which there are several types.

The simplest are \emph{potential fields}, $\B=\grad\Phi$ for some harmonic scalar potential $\Phi$, i.e., where $\nabla^2\Phi=0$. Then $\boldj=\mu^{-1}\Curl\B=\mathbf{0}$, by the standard vector field result that the curl of a gradient always vanishes. This trivially makes the Lorentz force zero. It can be shown that a potential field in a closed volume with $B_n=\hat\n\,\vdot\,\B$ specified on its boundary (unit normal $\hat\n$) is unique and has the minimum possible energy, making it globally stable. This is also true external to a closed surface such as the entire solar photosphere over which $B_n$ is prescribed if $B\to0$ as radius $r\to\infty$ is also assumed. Another name for a potential field is a \emph{current free field}.

The other way that a magnetic field can be force free is if the current is parallel to $\B$, since the cross product of parallel vectors always vanishes. Thus we may write $\curl\B=\alpha\,\B$ for some scalar function of position $\alpha$. However, $\alpha$ cannot be arbitrary. Recall that $\Div\B=0$ and div~curl is always zero, so $\B\,\vdot\,\grad\alpha=0$. But this is a directional derivative meaning that $\alpha$ must be constant along field lines. The particular case where all field lines have the same $\alpha$ is called a \emph{constant-$\alpha$ field}, for which (by a standard vector identity) $(\nabla^2+\alpha^2)\B=0$: the vector Helmholtz equation.

A simple example of a non-force-free MHS equilibrium, with gas pressure $p$ but no gravity, is the case of a non-uniform magnetic slab $\B=(0,0,B(x))$ for which total pressure $p(x)+B^2(x)/2\mu$ is uniform, since there are no net tension or pressure forces.

These insights can be invaluable when constructing force-free or MHS model equilibria for exploring waves. Of course, the Sun is a very dynamic place, and is certainly not in equilibrium. Nevertheless, there are large-scale magnetic structures (sunspot fields, coronal holes, systems of coronal loops, etc.) that persist for much longer than the wave-crossing times on which they might be expected to evolve. In this sense, a time-independent, static, magneto-atmosphere is a reasonable first approximation to what is at best a statistically steady equilibrium.

\subsection{Energy, Variational Principles and Stability}\label{sec:energy principle}
Of course, a MHS equilibrium -- where all forces balance -- may not be stable. A marble sitting on top of a smooth hill, or even a smooth saddle, may well be in equilibrium, but it is not stable, since there are tiny perturbations that could be made to its position or velocity that would see it run away down the hill. 

This may hold for MHS equilibria too. If an equilibrium is such that \emph{all} allowable infinitesimal perturbations to its state subsequently move back towards the equilibrium, the system is said to be \emph{linearly stable}. This is a necessary but not sufficient condition for \emph{global stability}, i.e., stability to all  perturbations, no matter how large.

To determine if a particular equilibrium is linearly stable, we must make a general linear perturbation to its state, for example by imposing a small non-zero velocity $\boldv$. This results in the Ferraro-Plumpton equation (\ref{FerraroPlumpton}) for MAG waves to be derived in Section \ref{sec:mag} from the MHD differential equations. The details do not concern us here, but the upshot is an equation of the form
\begin{equation}
    \pderivd{\boldv}{t} = -\G[\boldv]
\end{equation}
for a particular self-adjoint linear force operator $\G$ \citep[Section~6.2.3]{GoePoe04aa}. 

This linear perturbation equation may be Fourier transformed from the time $t$ to the frequency $\omega$ domain. In other words, we build 
${\boldv}({\x},t)$ from a weighted superposition of Fourier modes
\begin{equation}
{\V}(\x,\omega)
\exp\left[-i\omega t\right],
\end{equation}
where it is convenient to introduce the Lagrangian displacement of a parcel
of fluid from its equilibrium position defined by
\begin{equation}
\bxi(\x,\omega) \equiv \frac{i}{\omega} {\V}(\x,\omega)~.
\end{equation}

The Fourier-transformed Ferraro-Plumpton equation then takes the compact form
\begin{equation}
\omega^2 \bxi = \G[\bxi]~.
\end{equation}
The Energy Principle of \citet{FriRot60aa}  follows by taking the usual (Euclidean)
dot-product
of both sides of this equation with $\rho \bxi^*$, and integrating over $\x$.
This yields the familiar Rayleigh-Ritz formula for the square of the frequency:
\begin{equation}\label{RR}
\omega^2 = \frac{\langle \bxi^*, \G[\bxi] \rangle }{
\langle \bxi^*, \bxi \rangle}~,
\end{equation}
provided any surface-integrals may be neglected. If \emph{any} Lagrangian displacement
can be found for which the right-side of this equation is negative, then the 
equilibrium (encoded in the self-adjoint operator $\G$) is unstable. Otherwise,
this expression can be used to estimate the oscillation frequencies of an isolated
magnetostatic equilibrium, because the right-side of this equation is second-order in
the difference between the actual (usually unknown) Lagrangian displacement
$\bxi$, and any surrogate. 

Finally, we note that the actual Lagrangian displacement is an extremum of the action,
\begin{equation}\label{actionS}
S[\bxi] = \int \!d\x\, dt ~\rho(\x) \left(
\frac{\partial \bxi }{ \partial t}\,\vdot\,\frac{\partial \bxi }{ \partial t}
- \bxi\, \vdot \,\G[\bxi] \right)~,
\end{equation}
where here $\bxi$ has not been Fourier transformed and so is real; see, for example \cite{16ogilvie,16keppens1,16keppens2}. The Ferraro-Plumpton equation may be found variationally by setting $\delta S = 0$ \citep{FerPlu58aa,Tho83aa}.

\section{What are MHD Waves?}\label{sec:waves}
\subsection{Basic Equations}
Waves result from the interplay of fluid inertia and restoring forces. In this chapter, we restrict attention to \emph{linear waves}. Variables such as the density $\rho$ are written as $\rho_0+\rho_1+\rho_2+\cdots$. The subscript `0' indicates the equilibrium value and the subscript `1' denotes a small perturbation, $\rho_1 \ll \rho_0$. Terms of quadratic or higher order in the perturbation quantities are `neglected' in the sense that they usually serve as source terms which may be prescribed ab initio. If the equilibrium is stationary, then $\boldv=\boldv_1$ is intrinsically small, so $\rho_1\boldv$ for example is dropped, or perhaps accommodated in a source term.

The existence of magneto\-hydro\-dynamic waves was first predicted from theory over 80 years ago \citep{Alf42aa,Rus18mk}. They assume their simplest form in a homogeneous 
ideal-gas fluid
permeated by a uniform magnetic field. 
If this medium is perturbed from its stable equilibrium, the linearized MHD equations take the form
\begin{gather}
\pderiv{\rho_1}{t}+\rho_0\Div{\bf v}=0, \label{lincty} \\[4pt]
\rho_0\pderiv{{\bf v}}{t}=-\grad
p_1+\frac{1}{\mu}\left(\Curl\B_1\right)\vcross\,\B_0,  \label{linmmntm}\\[4pt]
\pderiv{p_1}{t}-c^2\pderiv{\rho_1}{t} =
(\gamma-1)\rho_0 T_0 \pderiv{\eta_1}{t} = 0,  \label{linenergy}\\[4pt]
\pderiv{\B_1}{t}=\curl\!\left({\bf v}\,\vcross\,\B_0\right),  \label{lininduct}\\[4pt]
\Div\B_1=0,
\end{gather}
where $c=\sqrt{\gamma p_0/\rho_0}$ is the adiabatic sound
speed, $T$ is the temperature (units: Kelvins), and
$\eta$ is the specific entropy (units: Joules kg$^{-1}$ Kelvin$^{-1}$). We do not explicitly write out the source terms in the conservation equations
for mass, momentum, and entropy as they may assume different forms based on a given
application. Without such terms or inhomogeneous initial or boundary conditions, these
equation have only the trivial solution. 

A very useful property of these equations is the existence of a wave-energy/flux
conservation law which is {\em quadratic} in the wave amplitudes:
\begin{equation}
    \pderiv{U_2}{t} + \Div{{\bf f}_2} = 0,
\end{equation}
where
\begin{gather}
    U_2 = \frac{1}{2} \rho_0 v^2 + \frac{p_1^2}{2\rho_0 c^2} + \frac{1}{2\mu} B_1^2~, \\
    {\bf f}_2 = p_1 {\bf v} + \frac{1}{\mu} ( \B_0 \cdot \B_1 ) {\bf v} - 
    \frac{1}{\mu}( {\bf v} \cdot \B_1) \B_0~.
\end{gather}
The wave energy-flux, ${\bf f}_2$, permits one to assess the potential for MHD waves to
heat a stellar atmosphere. Notice that the wave energy-density is distributed across
kinetic, thermal and magnetic reservoirs. The wave-energy/flux conservation law also applies to pure acoustic waves ($\B_0 = \mathbf{0}$).

Differentiating Equation (\ref{linmmntm}) with respect to time and eliminating $\rho_1$,
$p_1$, and $\B_1$ in favour of $\bf v$ using the remaining equations, we find
\begin{equation}
\pderivd{\boldv}{t}=c^2\grad(\Div{\boldv})
+\frac{1}{\mu\rho_0} \left(\curl\left(\curl\left({\boldv}\vcross\B_0\right)\right)\right)\!\vcross\B_0,
\end{equation}
which is beginning to look more like the familiar wave equation, though the magnetic term is at first perplexing. It is an {\em anisotropic, vector,} wave equation.

On the other hand, the coefficients that appear in this equation are by construction
all constants. The equation is invariant under translations in space and time.
It is invariant under reflections in time, and arbitrary rotations about the equilibrium
magnetic field. Therefore, it may be solved by standard Fourier transform methods. 

The idea is to build ${\boldv}({\x},t)$ from a weighted superposition of Fourier modes
\begin{equation}
{\V}(\k,\omega)
\exp\left[i({\k\,\vdot\, \x}-\omega\, t)\right],
\end{equation}
where $\V$ is a velocity amplitude vector, which satisfies
the three algebraic equations
\begin{equation}\label{fullwv}
 \begin{split}
\omega^2{\bf V} &=c^2{\bf k}({\k\,\vdot \V}) +
\frac{1}{\mu\rho_0} \left({\k}\,\vcross\left({\k}\,\vcross\left({\bf
V}\vcross\B_0\right)\right)\right)\vcross\B_0\\
&= c^2\k (\k\vdot\V) \\ &\qquad +
\left[(\a\vdot\k)^2\V-(\k\vdot\a)(\k\vdot\V)\a
-(\k\vdot\a)(\a\vdot\V)\k+a^2(\k\vdot\V)\k\right].
 \end{split}
\end{equation}
The Alfv\'en velocity $\a=\B_0/\sqrt{\mu\rho_0}$ and Alfv\'en speed $a=|\a|$ have been introduced.\footnote{Other common notations for the sound and Alfv\'en speeds are $c_s$ and $v_A$ respectively.} Again, we have omitted an inhomogeneous source vector obtained from the
Fourier transform of the source terms in the original PDEs.

The latter form suggests projecting in the $\hat\k$, $\hat\a$ and $\widehat{\k\vcross\a}$ directions, where the hat indicates a unit vector. This returns a matrix equation
\begin{equation}\label{matDisp}
    \begin{pmatrix}
              \omega^2-(a^2+c^2)k^2 & a^2 k \kpar & 0 \\
              -c^2 k \kpar & \omega^2            & 0 \\
              0           & 0            & \omega^2-a^2 \kpar^2
    \end{pmatrix}
    \begin{pmatrix}
      \hat\k\,\vdot\,\V\\ \hat\a\,\vdot\,\V\\ \widehat{\k\,\vcross\,\a}\,\vdot\V
    \end{pmatrix}
    = \text{sources},
\end{equation}
where $\kpar=\hat\a\,\vdot\,\k= k\cos\alpha$ is the wavevector component parallel to the magnetic field and $\alpha$ is the angle between $\k$ and $\B_0$. The `sources' on the
right side of this equation is a prescribed 3-vector which generally depends upon both
$\k$ and $\omega$. Regarding $\k$ and $\omega$ as 4 independent complex variables, one
now inverts the 
$3\times3$ matrix to solve for the three linearly-independent components of $\V$. This
solution is then inverse Fourier transformed to find the desired ${\boldv}({\x},t)$. In the absence of sources, the $3\times3$ matrix must be singular for there to be non-trivial solutions.

\subsection{Dispersion Relation}\label{sec:dispersion}
An essential component of this process is computing the determinant of the 
$3\times3$ matrix, which is usually called the propagator or the dispersion matrix. The integrand of the
inverse Fourier transforms has singularities, usually in the form of isolated poles,
where the determinant vanishes. 
Setting the determinant to zero yields the \emph{dispersion relation}
\begin{equation}\label{MHD disp}
    \left(\omega^2-a^2\kpar^2\right)\left(\omega^4-(a^2+c^2)k^2\omega^2+a^2c^2k^2\kpar^2\right)=0.
\end{equation}
The zeros (simple poles) provide distinct plane-wave solutions (also called modes)
with resultant phase speeds
\begin{equation}
    \frac{\omega}{k}=\pm a\cos\alpha,
\end{equation}
the Alfv\'en wave, and
\begin{equation}
\begin{split}
\frac{\omega}{k}&=\pm\left[\half(a^2+c^2)\pm
\half\sqrt{(a^2+c^2)^2-4a^2c^2 \cos^2\alpha}\right]^{1/2}\\[4pt]
&=\pm\left[\half(a^2+c^2)\pm
\half\sqrt{a^4+c^4-2a^2c^2 \cos 2\alpha}\right]^{1/2},
\end{split}
\end{equation}
the fast ($+$ sign inside the square brackets) and slow ($-$ sign inside the brackets) waves. It is easily shown that $(\omega/k)_{\mbox{\scriptsize slow}}\le \min(a,c)\le\max(a,c)\le(\omega/k)_{\mbox{\scriptsize fast}}\le\sqrt{a^2+c^2}$, with the last relation being an equality only if $\cos\alpha=0$.

Notice that the dispersion relation is third order in $\omega^2$ and $\kpar^2$, but 
it is only
second order in $\kperp^2=k^2-\kpar^2$. This results from the anisotropy induced by the equilibrium
magnetic field. It has some important consequences for horizontal magnetic fields and
waves in magnetic flux tubes. Only even powers of frequencies and wavenumbers are
present because of the invariance of the equations under time reflection. When dissipation
is present, this symmetry is broken. Finally, it is worth remembering that the
Fourier transform of the source vector may also have singularities. The presence of 
square-roots invariably produces branch-point singularities, which must be joined
in some fashion by branch cuts to ensure that quantities are single-valued. These will contribute to the weighted superposition, via contour integrals rather than residues.

\begin{figure}[htbp]
\begin{center}
\includegraphics[width=.9\textwidth]{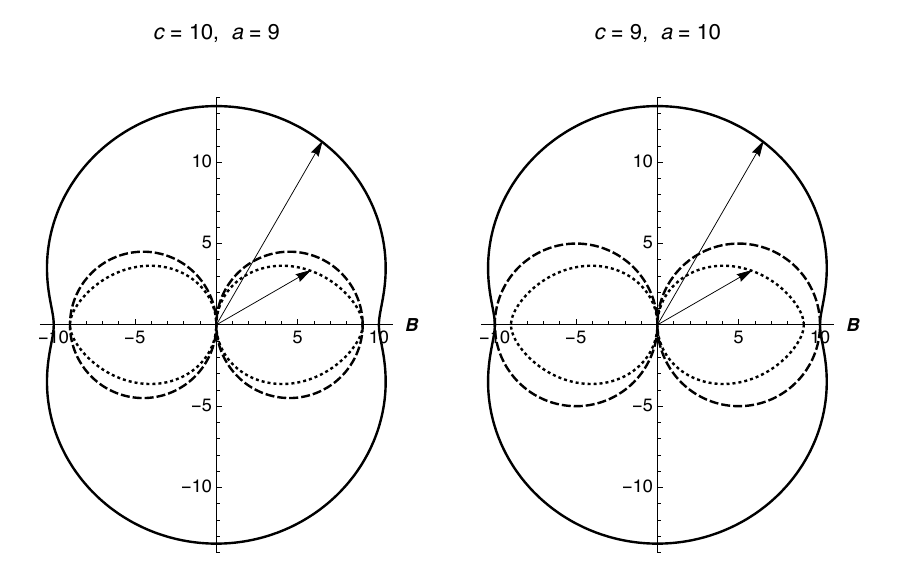}
\caption{Magneto-acoustic polar diagram showing the phase speeds of both the
fast (full curves) and slow waves (dotted) propagating at angle $\alpha$ to $\B$, which is aligned with the $x$-direction. In the left diagram we
assume $c=10$ and $a=9$, i.e., $a>c$, whereas on the right $c=9$ and $a=10$. 
The Alfv\'en locus is included for comparison (dashed) in each case.}
\label{fig:fried_ph}
\end{center}
\end{figure}

Figure \ref{fig:fried_ph} illustrates the differing phase speeds $\omega/k$ of the three wave types. With $\B_0$ oriented in the $x$-direction and $\k$ in the $x$--$y$ plane, the arrows are examples of $(\omega/k)\hat\k$ ending on one or other of the phase-speed loci. Their lengths indicate the phase speed of the wave type in the selected direction.

The simplest of the three wave types in the Alfv\'en wave, driven solely by magnetic tension, which Equation (\ref{matDisp}) shows is both incompressive $\k\,\vdot\,\V=0$, i.e., $\Div\boldv=0$, and transverse to the magnetic field, $\B_0\,\vdot\,\V=0$. The fast and slow waves are driven by a combination of plasma and magnetic pressure and magnetic tension.

Letting $\V=(U,V,W)^T$ in Cartesian coordinates, with $\B_0$ arbitrarily aligned with the $x$-direction and $\k$ lying in the $x$--$y$ plane, the matrix equation can be rearranged in eigenvalue form $A\V=(\omega/k)^2\V$, with
\begin{equation}\label{Amat}
   A= \begin{pmatrix}
      c^2\cos^2\alpha & c^2\sin\alpha\cos\alpha & 0 \\
      c^2\sin\alpha\cos\alpha & a^2+c^2\sin^2\alpha & 0 \\
      0 & 0 & a^2 \cos^2\alpha
    \end{pmatrix}.
\end{equation}
Being symmetric and positive definite, the eigenvalues $(\omega/k)^2$ of $A$ are necessarily real and positive, as already seen above, but also its eigenvectors $\V$ are orthogonal. That is, for a given direction $\alpha$ of $\k$, the velocity polarizations of the three wave types are mutually orthogonal.

Often we are confronted with situations in which there are great disparities between
the magnitudes of the sound speed and the Alfv\'en speed. Cold plasmas obtain in the 
limit $c \rightarrow 0^+$. In this limit, the slow mode ceases to propagate; the fast mode
propagates isotropically. The incompressive limit (i.e., $\Div \boldv \rightarrow 0$)
obtains in the opposite extreme where $c \rightarrow \infty$. The fast mode assumes
unphysical phase speeds and is discarded. 
The slow mode remains, but it is unable to propagate
perpendicular to the magnetic field. In both limits, the Alfv\'en mode is unaffected.

\subsection{Phase and Group Velocities}\label{ph gr}
In multiple dimensions, the \emph{phase velocity} 
\begin{equation}\label{vph}
    \boldv_\text{ph}=\frac{\omega}{k}\,\hat\k,
\end{equation}
where $k=|\k|$ is the wavenumber, indicates the speed and direction that the peaks and troughs of the wave are travelling. On the other hand, the \emph{group velocity}
\begin{equation}\label{vgr}
    \boldv_\text{gr} = \pderiv{\omega}{\k} = \left(\pderiv{\omega}{k_x},\,\pderiv{\omega}{k_y},\,\pderiv{\omega}{k_z}\right)
\end{equation}
represents the speed and direction of energy propagation.

Strictly, the concept of `group' velocity only makes sense in a Fourier superposition of frequencies and wavenumbers, though it even applies in nascent form with just two infinitesimally separated 1D monochromatic waves $(\omega+\Delta\omega,k+\Delta k)$ and $(\omega-\Delta\omega,k-\Delta k)$, where the envelope of their beating travels at speed $\Delta\omega/\Delta k$:
\[
e^{i((k+\Delta k)x-(\omega+\Delta\omega)t)} +
 e^{i((k-\Delta k)x-(\omega-\Delta\omega)t)}    =
 2 e^{i(kx-\omega t)} \cos(\Delta k\,x-\Delta\omega\, t).
\]
See \citet[][Sections~11.4--11.6]{Whi74aa} for a more nuanced discussion. In practice, no MHD wave is truly monochromatic, so we will persist with loosely referring to the group velocity of a single wave. 

In calculating group velocity, we differentiate the dispersion relation $\omega(\k)=0$, either explicitly or implicitly. For the Alfv\'en wave $\omega^2=a^2\kpar^2$, this gives
\begin{equation}\label{vgrA}
    \boldv_\text{gr,A}=a \,\hat\B_0.
\end{equation}
That is, irrespective of the direction the wave pattern appears to be travelling in, the wave energy is actually propagating directly along the magnetic field lines at the Alfv\'en speed. Alfv\'en wave energy does not cross field lines. This makes sense when we recall that Alfv\'en waves are driven by magnetic tension alone, so they are like waves on a taut string.

\begin{figure}[htb]
\begin{center}
    \includegraphics[width=.45\textwidth]{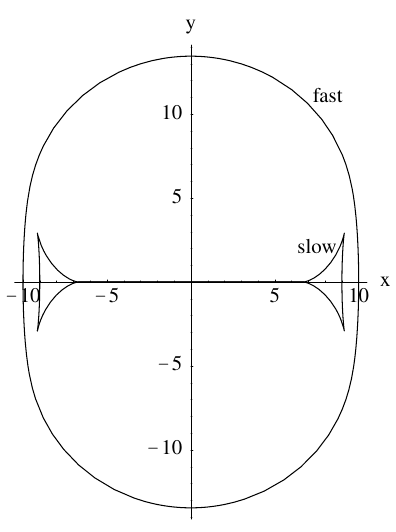}\hfill
    \includegraphics[width=.45\textwidth]{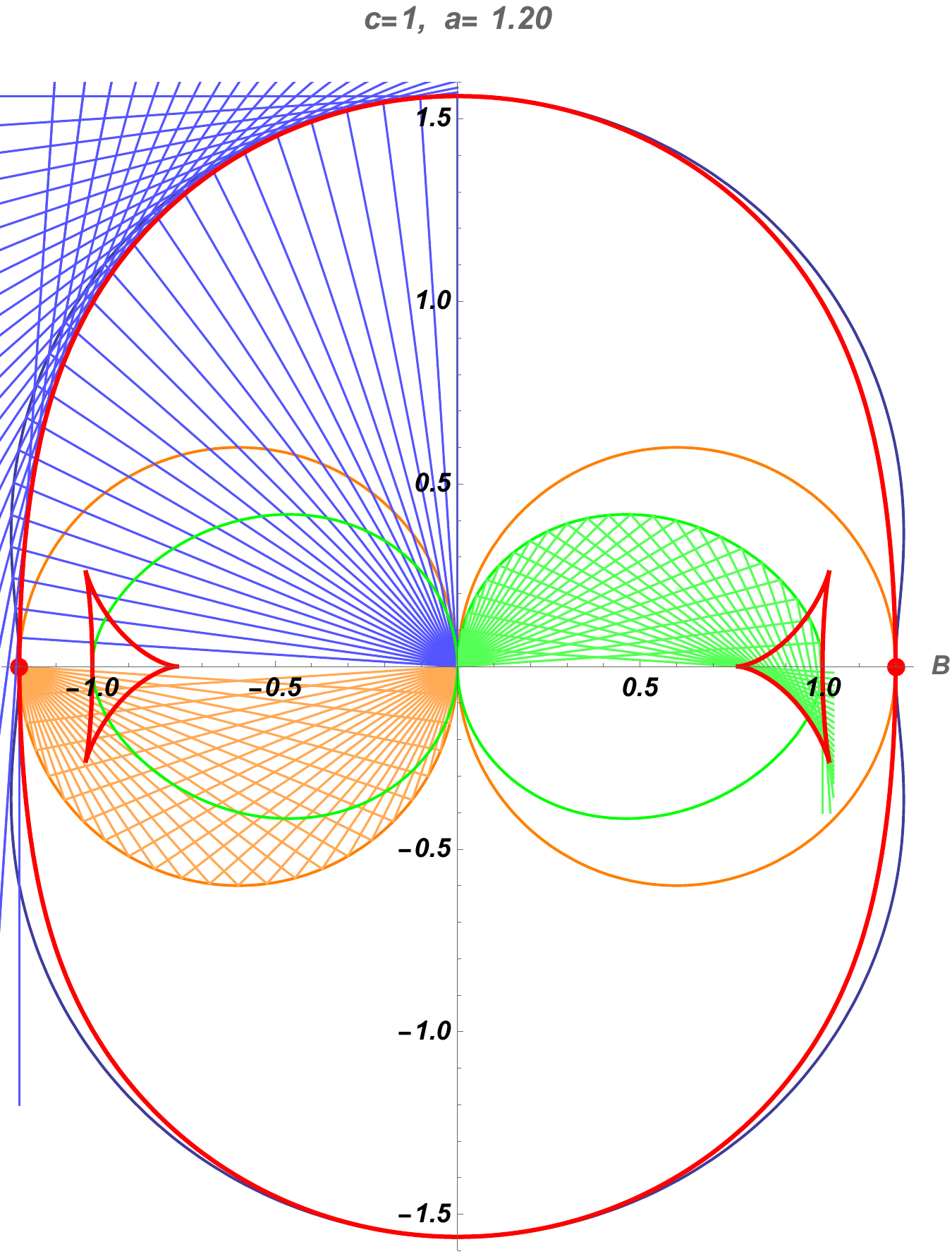}
\caption{Left: Polar diagram illustrating the shape of fast and slow MHD pulses, evolving according to their group velocities, assuming the magnetic field is oriented in the $x$-direction.
The fast wave expands in all directions, whilst the slow wave is
restricted to a narrow sector about the field direction. As in the left frame
of Figure \ref{fig:fried_ph}, $c=10$, $a=9$ is assumed here, for which the cusp speed is $\cT=6.69$. Right: geometrical construction of a group speed polar diagram. See text for details.}
\label{fig:fried_gr}
\end{center}
\end{figure}

For the magneto-acoustic waves, the group velocity is most easily calculated by introducing the notation $D(\omega,\k)$ for their dispersion function, the second factor in Equation (\ref{MHD disp}), and differentiating implicitly to get $(\partial D/\partial\omega)\,\partial\omega/\partial\k+\partial D/\partial\k=0$ by the chain rule, or
\begin{equation}
\pderiv{\omega}{\k}=-\frac{\partial D/\partial\k}{\partial D/\partial\omega}.
\end{equation}
Specifically, the group speed parallel to the magnetic field is then
\begin{equation}\label{vgr par}
\pderiv{\omega}{\kp}=\frac{(\omega^4-a^2c^2k^4)\kp}{\omega k^2[2\omega^2
-(a^2+c^2)k^2]},
\end{equation}
and the perpendicular speed is
\begin{equation}\label{vgr perp}
\pderiv{\omega}{\kperp}=\frac{\omega^3\kperp}{k^2[2\omega^2-(a^2+c^2)k^2]},
\end{equation}
where $\kperp$ is the component of $\k$ perpendicular to $\B_0$.
There is no energy propagation in the direction perpendicular to both $\B_0$ and $\k$.

A little algebra reveals that the group velocity of a fast wave propagating in the $\hat\B_0$ direction is the same as its phase velocity, $\max(a,c)\,\hat\B_0$, and similarly when propagating perpendicularly it is $\sqrt{a^2+c^2}\,\hat\k$. For intermediate directions, the group and phase speeds differ, as is clear from Figure \ref{fig:fried_gr}.

The slow wave is even less isotropic. For parallel propagation its group velocity is $\cT\,\hat\B_0$, where $\cT=a\,c/\sqrt{a^2+c^2}$ is the so-called cusp speed, for reasons obvious from Figure \ref{fig:fried_gr}.

Figure \ref{fig:fried_gr} (left panel) represents the shapes of the disturbed regions resulting from an MHD pulse emitted from the origin. The outermost wave front propagates at the group velocity. Depending on whether it was a fast wave pulse or a slow wave pulse, the two displayed shapes evolve. The shapes of these regions, especially for the slow wave, may be made clearer by geometric construction (right panel). Starting with the phase velocity polar diagram (blue for fast, orange for Alfv\'en, green for slow), draw a straight line from the origin to the phase locus. This is $\k$. From there, draw the perpendicular from that line; this represents the constant phase surfaces (perpendicular to $\k$). Do this for a large number of directions. You will notice, that these surfaces form envelopes. They are the group-velocity loci. The mysterious slow wave cusp is now easier to understand. Notice in particular that the \emph{lower} half of the slow wave cusp is constructed from wave vectors oriented in the \emph{upward} directions (green lines as drawn). That is, the phase and group velocities are directed on opposite sides of the $\hat\B_0$ direction, which may be verified algebraically using Equation (\ref{vgr perp}) since the denominator is negative for the slow wave.

For the Alfv\'en wave, the group locus degenerates to a single point (the red dot) due to the well-known property of circles that an inscribed triangle with one side forming the diameter is a right-angled triangle.

\begin{figure}[htb]
    \centering
    \includegraphics[width=\textwidth]{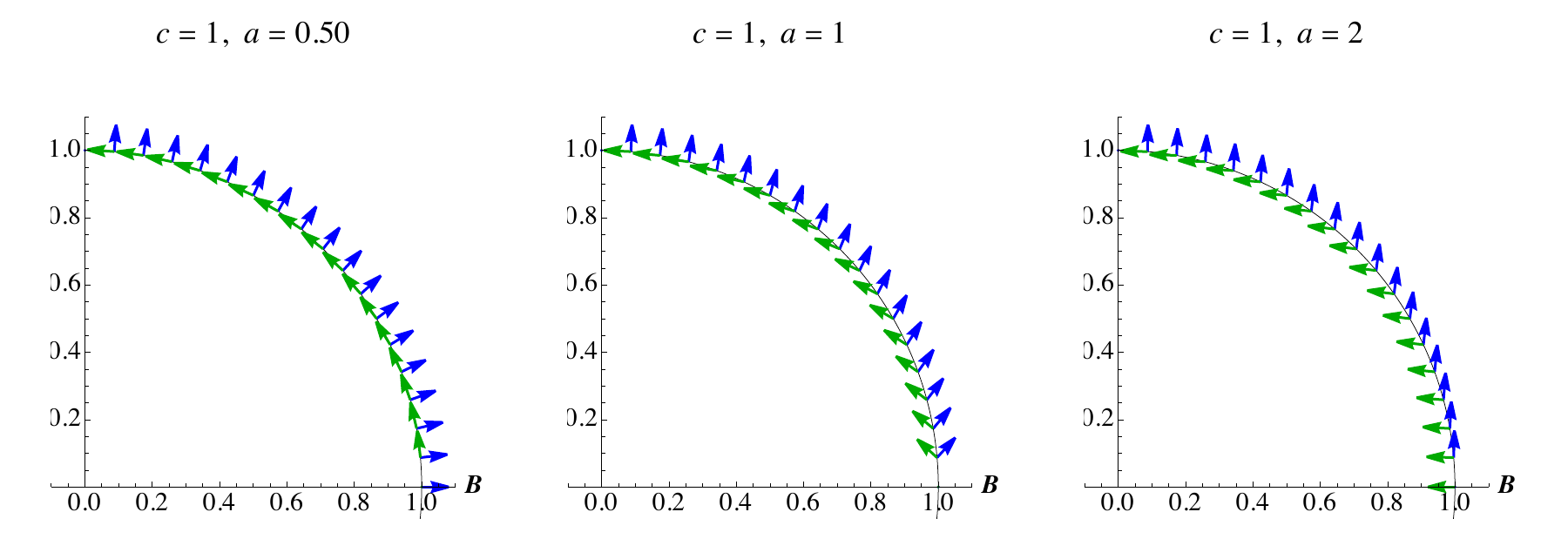}
    \caption{Velocity polarization directions of the slow (green) and fast (blue) waves for $a/c=0.5$ (left), 1 (centre) and 2 (right). The positions of the arrows correspond to the directions of the wave vector $\k$.}
    \label{fig:polarizations}
\end{figure}

\subsection{Wave Polarizations}\label{sec:polariz}
It remains to emphasise the velocity polarizations of the three wave types. These are the eigenvectors $\V$ of the real symmetric matrix $A$ defined in Equation (\ref{Amat}), and so they are necessarily orthogonal. The Alfv\'en wave is the simplest; it is polarized in the direction $\k\vcross\B_0$, perpendicular to both $\k$ and $\B_0$. The fast and slow wave velocities therefore both lie in the plane spanned by those two vectors. Their directions are illustrated in Figure \ref{fig:polarizations} for three ratios of the Alfv\'en to sound speeds. It is easily verified that the fast-wave polarization is longitudinal (i.e., in the $\k$ direction) for $a\ll c$, where it is essentially just the sound wave. The slow wave is therefore transverse to $\k$ in this limit. On the other hand, for $a\gg c$, the slow wave is just a field-guided sound wave restricted to have its velocity along $\B_0$, and so the fast wave is asymptotically transverse to the magnetic field. These considerations are very useful in understanding MHD waves in stratified atmospheres, where polarizations vary with height.

\section{Waves in Stratified Atmospheres}\label{sec:waves strat}
Having introduced the basic MHD waves in a uniform plasma in Section \ref{sec:waves}, we now turn to the main topic of this chapter. How are these modified, or indeed do they even exist, in a continuously varying or stratified atmosphere like the solar chromosphere? 

\begin{figure}[htb]
    \centering
    \includegraphics{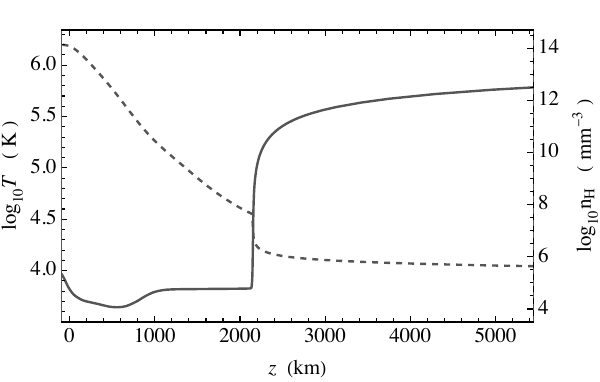}
    \caption{The temperature (full curve, left axis) and total hydrogen number density (dashed curve, right axis) plotted logarithmically as functions of height $z$ (km) in the mean quiet Sun Model C7 of \citet{AvrLoe08aa}. The transition region around 2200 km presents a formidable barrier to waves attempting to enter the corona from below.}
    \label{fig:TnH}
\end{figure}

\begin{figure}[htbp]
\begin{center}
\includegraphics[width=.95\textwidth]{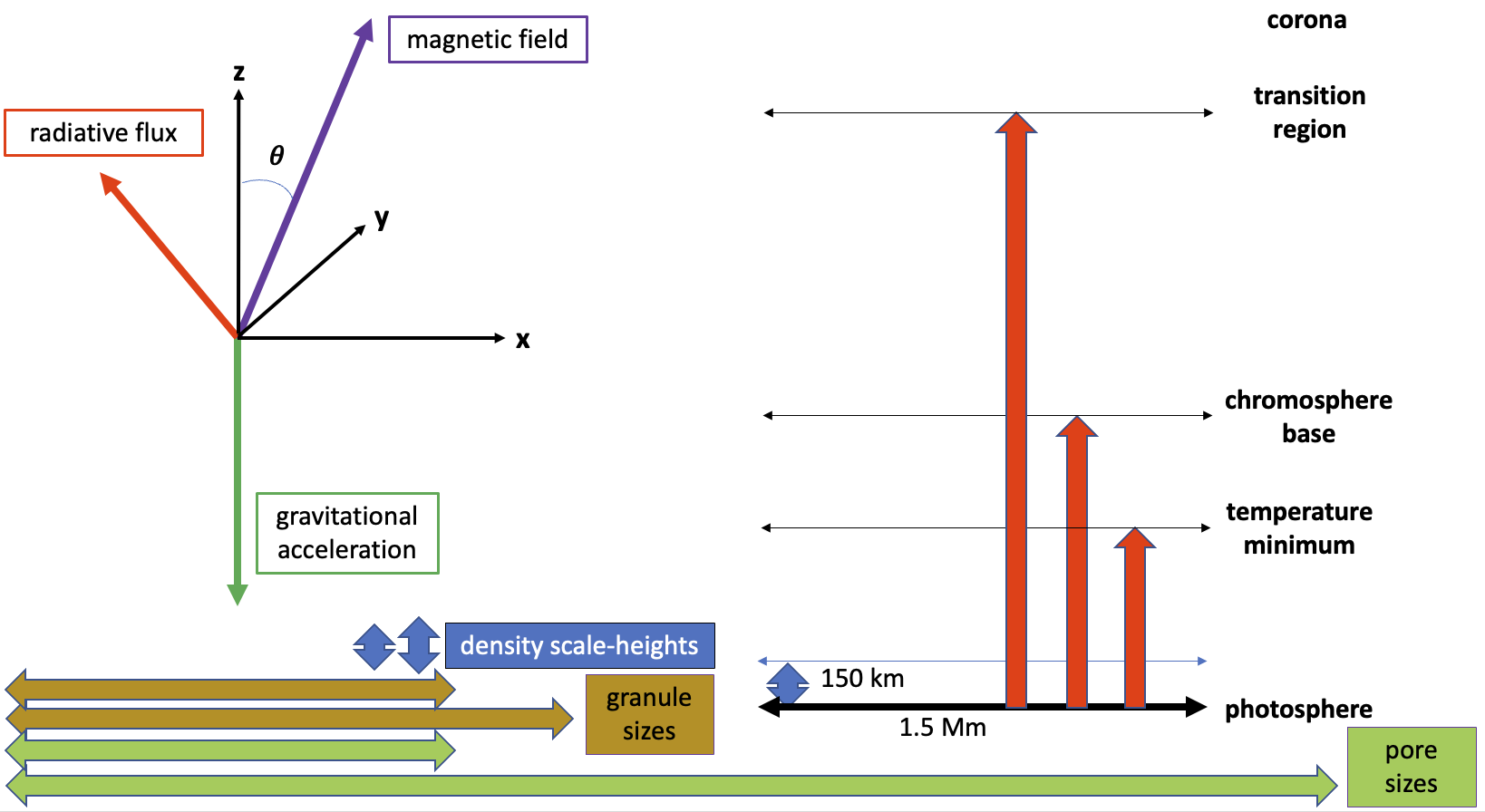}
\caption{Schematic illustrating representative widths and heights in the solar atmosphere. The shorter and longer arrows corresponding to granules and pores indicate size ranges. The axes and arrows at top left indicate symmetry-breaking features that may cause MHD wave modes to interact.}
\label{fig:WidthsAndHeights}
\end{center}
\end{figure}

Typically, the (mean) chromosphere is around 14 density scale-heights thick (see Figure~\ref{fig:TnH}), so gravitational stratification exerts a powerful influence on any waves. Figure \ref{fig:WidthsAndHeights} illustrates the extent to which vertical stratification dominates the low atmosphere. Even `tiny' features such as pores and granules are many times wider than their density scale height at the photosphere, making gravitational stratification the dominant feature affecting wave propagation. Stratified atmospheres will be our main topic for the rest of this chapter.

\subsection{Acoustic-Gravity Waves and the WKBJ Method}\label{sec:AGW}
Before moving to MHD waves in stratified atmospheres, it is useful to explore the effects of stratification without the magnetic field. The two remaining restoring forces are buoyancy and gas pressure via compression. 

Following \citet{DeuGou84aa}, introducing $\psi=\rho^{1/2}c^2\Div\bxi$ based on a method of \citet{Lam32aa}, the linearized oscillation equations can be reduced to a single second-order normal-form ordinary differential equation (ODE) for the amplitude $\Psi(z)$, where $\psi(x,y,z,t)=\Psi(z)\exp[i(k_xx+k_yy-\omega t)]$:
\begin{equation}
\derivd{\Psi}{z}=Q(z) \Psi,   \label{AGeqn}
\end{equation}
where
\begin{equation}
    Q(z)=k_h^2-\frac{\omega^2 -\omega_c^2}{c^2} -\frac{N^2}{\omega^2}k_h^2, \label{Qwkb}
\end{equation}
and
\begin{equation}
\omega_c^2 = \frac{c^2}{4H^2}\left(1-2H'\right)  \label{omegac}
\end{equation}
is the square of the (or more properly \emph{an}) acoustic cutoff frequency, $H_\rho = H=-\rho/\rho'$ is the density scale height, $N^2=g/H-g^2/c^2$ is the square of the {\bv} (buoyancy) frequency, and $k_h^2=k_x^2+k_y^2$ is the squared horizontal wavenumber.

Here $' = d/dz$, and all equilibrium quantities may generally depend upon the vertical
coordinate, $z$. The translational invariance in time and the two horizontal coordinates have been exploited to reduce the PDEs to ODEs in $z$. In place of the dispersion relation,
or the inversion of a $3\times3$ matrix, one is faced with solving a linear 
second-order ODE with three (possibly) complex parameters $\omega, k_x, k_y$. When $\Psi$ is determined, 
$\psi(x,y,z,t)$ is again obtained by inverting three Fourier transforms. As before, we
have for convenience suppressed a source term, which depends upon the three complex Fourier 
variables and $z$, in Equation (\ref{AGeqn}), 
to avoid the trivial solution $\Psi = 0$. The solution
of the resulting inhomogeneous equation must necessarily include the Wronskian of the two
linearly-independent solutions.

A wave-energy/flux conservation law also holds for acoustic-gravity waves. The appropriate
expressions are now
\begin{gather}
    U_2 = \frac{1}{2} \rho_0 v^2 + \frac{p_1^2}{2\rho_0 c^2} + 
    \frac{\rho_0 N^2}{2} \left( \frac{\eta_1}{\eta_0'} \right)^2 ,\\[4pt]
    {\bf f}_2 = p_1 {\bf v}.
\end{gather}
Notice that acoustic-gravity waves generally have non-zero (Eulerian) entropy 
fluctuations. The Lagrangian (comoving) entropy fluctuation is exactly zero. 
For an isentropic stratified atmosphere (i.e., $\eta_0' = N^2 = 0$), the third term (thermobaric energy density)
in the expression for the energy density is absent: both Lagrangian and Eulerian
entropy fluctuations vanish. 

\begin{figure}[htbp]
\begin{center}
\includegraphics[width=.6\textwidth]{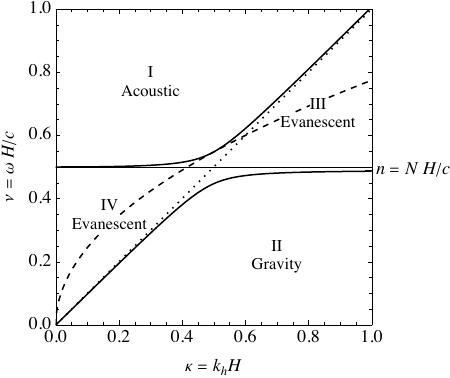}
\caption{Regions of the 
\hbox{$k_x-\omega$} plane inhabited by modified
acoustic, modified gravity, and evanescent waves. This diagram is calculated for an isothermal atmosphere, but will be similar for more general cases. Region I corresponds to vertically propagating acoustic waves, Region II to propagating gravity waves, and Regions III and IV to evanescent waves. The dotted line is the Lamb wave $\nu=\kappa$ and the dashed curve is the f-mode $\nu^2=\kappa/\gamma$. For the case $\gamma=5/3$.}
\label{fig:propdiag}
\end{center}
\end{figure}

In the case of an isothermal atmosphere, where $c$, $H=c^2/(\gamma g)$ and $N^2=(\gamma-1)g^2/c^2$ are all constant, Equation (\ref{AGeqn}) may be solved exactly in terms of elementary functions, $\Psi_{\mbox{\scriptsize iso}}\propto\exp(\pm i k_z z)$, where: 
\begin{equation}
    k_z^2=\frac{\omega^2 -\omega_c^2}{c^2} +\frac{N^2}{\omega^2}k_h^2-k_h^2.  \label{acGravDisp}
\end{equation}
Waves are therefore travelling vertically (or standing) if $k_z^2>0$ and evanescent if $k_z^2<0$. Arbitrarily setting $k_y=0$, $k_h=k_x$, Figure \ref{fig:propdiag} partitions the $k_x$--$\omega$ plane into travelling (Region I and II) and evanescent (III and IV) regions. A dimensionless wavenumber $\kappa=k_x H$ and frequency $\nu=\omega H/c$ are used. The dimensionless acoustic cutoff frequency in these units is exactly $\half$. Region I, which is entirely above this cutoff, hosts acoustic waves somewhat modified by gravity, progressively less so as frequency increases. Region II, which is entirely below both the cutoff and {\bv} frequencies, hosts internal gravity waves somewhat modified by acoustic effects.

A semi-infinite isothermal atmosphere provides a boundary along which an evanescent 
(Region III and IV) acoustic-gravity wave may propagate. The Lamb wave ($\omega^2=k_h^2 c^2$)
lives in an atmosphere with a rigid lower boundary, like the Earth's atmosphere. The 
surface gravity wave, or f-mode ($\omega^2=gk_h$), lives in an atmosphere with a stress-free
upper boundary, like a stellar convective envelope.

Returning to the more general (non-isothermal) case, exact solutions may, or may not, 
exist in terms of standard tabulated special functions. As this equation is equivalent
to the standard one-dimensional time-independent Schr\"odinger equation with a 
potential $Q(z)$, there exist tabulations of $Q$s for which the equation yields familiar
special functions. Their Wronskians and dispersion relations are readily computed.

The analytic properties of $Q(z)$, with $z$ taken to be a complex variable, determine
the singular points of the ODE, and enable one to classify the ODE and obtain Frobenius
and asymptotic expansions valid in the neighborhoods of the singular points. Numerical
methods may be employed between the singular points to connect the linearly-independent
solutions around the singular points.

The eikonal or WKBJ method\footnote{Named after Wentzel, Kramers and Brillouin who popularized it independently in  the context of quantum mechanics in 1926, and sometimes Jeffreys who contributed three years earlier without being widely recognised. Most commonly, the method is called WKB, but some authors append or even prepend the J.} \citep{BenOrs78aa} yields a high-frequency asymptotic-solution. For example, if $Q(z)<0$ throughout, corresponding to Region I or II in the isothermal case, an upward ($+$ sign) or downward ($-$ sign) travelling wave would satisfy
\begin{equation}
    \Psi \sim C \left[-Q(z)\right]^{-1/4} \exp\left[\pm i\!\int^z \sqrt{-Q(z')}\,dz'\right]
\end{equation}
as $|Q|\, H^2\to\infty$, which in effect can mean $\omega\to\infty$ in Region I or $\omega\to0$ in Region II. Similarly if $Q(z)>0$ throughout, the upwardly evanescent wave is represented asymptotically by
\begin{equation}
    \Psi \sim C\, Q(z)^{-1/4} \exp\left[-\int^z \sqrt{Q(z')}\,dz'\right].
\end{equation}

These eikonal asymptotic formulae do not apply near a turning point $z_t$ (corresponding to the boundaries of Regions I and II in figure \ref{fig:propdiag}) at which $Q(z)=0$ with $Q'(z_t)>0$. However, the distinguishing feature of the WKBJ method that elevates it beyond the simple eikonal method is that a matching across $z_t$ can be developed that is valid uniformly across the domain \citep[the Langer solution;][]{BenOrs78aa}:
\begin{equation}
    \Psi(z) \sim C\,\left(\frac{3}{2}S_0(z)\right)^{1/6} Q(z)^{-1/4} \Ai\left[\left(\frac{3}{2}S_0(z)\right)^{2/3}\right], \label{PsiLanger}
\end{equation}
where $S_0(z) = \int_{z_t}^z \sqrt{Q(z')\,dz'}$, $C$ is a normalization constant and $\Ai$ is the Airy function of the first kind. This is useful in modelling the effect of the acoustic cutoff in confining low frequency ($\,\lesssim 5$ mHz) solar p-modes. (It is assumed that $z_t$ is a simple isolated zero of $Q$.) Care must be taken with Equation (\ref{PsiLanger}) because of the fractional powers and $S_0$ being imaginary on $z<z_t$. Specifically, on $z<z_t$ one must select $\arg S_0=3\pi/2$, so that $\arg S_0^{1/6}=\pi/4$ and $\arg S_0^{2/3}=\pi$. The principal real positive root applies on $z>z_t$.

Of course, the exact location of the turning point of a wave, identified here with $Q=0$ and hence with an inflection point $\Psi''=0$, depends crucially on the choice of dependent and independent variables used to express the wave equation. Different choices lead to different expressions for the cutoff frequency \citep{SchFle98aa,SchFle03aa}, though the {\bv} frequency is not affected.\label{omc} This makes it difficult to interpret observations in terms of a height-dependent cutoff frequency. In addition, formulae such as (\ref{omegac}) that involve higher derivatives, in this case the second $z$-derivative of the density, produce very spiky cutoff frequency profiles when applied to tabulated empirical atmosphere models such as the widely used Model C of \citet{VerAvrLoe81aa} (the VAL C model). This complicates both interpretation and numerical modelling using, in particular, ray theory.

The acoustic cutoff, and how it is modified by magnetic field, plays an important role in the propagation of waves through the solar chromosphere, as explored in Section \ref{sec:exact2D}.

\subsection{MAG Waves}\label{sec:mag}
The generalization of Section \ref{sec:waves} to an arbitrarily-stratified
stationary magneto-atmosphere is straightforward. 
\begin{gather}
\pderiv{\rho_1}{t}+\Div{\rho_0 \bf v}=0, \label{linctyg} \\[4pt]
\rho_0\pderiv{{\bf v}}{t}=-\grad
p_1+\rho_1{\mathbf g}_0 +
\frac{1}{\mu}\left(\Curl\B_0\right)\vcross\,\B_1 +
\frac{1}{\mu}\left(\Curl\B_1\right)\vcross\,\B_0,  
\label{linmmntmg}\\[4pt]
\Deriv{p_1}{t}-c^2\Deriv{\rho_1}{t}= (\gamma-1) \rho_0 T_0 
\Deriv{\eta_1}{t} = 0,
\label{linenergyg}\\[4pt]
\pderiv{\B_1}{t}=\curl\left(\boldv\,\vcross\B_0\right),  \label{lininductg}\\[4pt]
\Div\B_1=0.
\end{gather}
By assumption, the equilibrium quantities satisfy the three MHS constraints:
\begin{equation}
    0=-\grad p_0 + \rho_0\,{\mathbf g}_0 +
    \frac{1}{\mu}\left(\Curl\B_0\right)\vcross\,\B_0,\label{MHS}
\end{equation}
and $\Div \B_0 = 0$.

The neglect of the term $\rho_0{\mathbf g}_1$ in the linearized equations is known
as the Cowling approximation. It is extremely accurate in most situations. 
Accordingly, we may simply set ${\mathbf g}_0 = {\mathbf g}$. The adiabatic sound
speed $c$ (and the Alfv\'en speed $a$) is also derived from the equilibrium pressure, density, and ratio
of specific heats. It may depend upon all three spatial coordinates. 
All dissipative (i.e., non-ideal) terms have been omitted from these equations.

As before, it proves possible to reduce these to a {\em single} vector wave-equation for
$\bf v$:
\begin{multline}
\pderivd{\boldv}{t}= \frac{1}{\rho_0} \grad(
\rho_0 c^2\Div{\boldv}) + \grad(\boldv \cdot {\mathbf g}) -
{\mathbf g}(\Div \boldv)
+\frac{1}{\mu\rho_0}
\left(\curl\left(\curl\left({\boldv}\vcross\B_0\right)\right)\right)\!\vcross\B_0 \\
+ \frac{1}{\mu\rho_0} \grad \left( \boldv \cdot \left( \curl \B_0 \right)
\right) +
\frac{1}{\mu\rho_0} \left( \left( \curl \B_0 \right) \vcross \left( \curl
\left( \boldv \vcross \B_0 \right) \right) \right)
\equiv - \G[\boldv], \label{FerraroPlumpton}
\end{multline}
a result which was first derived by \cite{FerPlu58aa}. This result is
exact given our assumptions. It describes not only MAG waves that propagate through
stable equilibria, but it will also capture the ideal instabilities (see Section~\ref{sec:energy principle}).

A wave-energy/flux conservation law may again be deduced from this set of linearized
equations. It is the obvious hybrid
obtained by combining the previous results for the acoustic-gravity and MHD waves.

\subsection{Exact 2D MHD Solutions and their Mixed Properties} \label{sec:exact2D}
Exact solutions in any form of modelling have value beyond their strict applicability to reality. They help us understand processes and possibilities, and also provide 
rigorous tests for numerical schemes. The combination of all three restoring forces -- gas pressure, buoyancy and Lorentz force -- in a non-trivial exact solution obtained by \citet{ZhuDzh84aa} was therefore a most welcome, if unexpected, innovation. 

Their model is of ideal MHD waves in a plane-stratified isothermal atmosphere with uniform non-horizontal magnetic field in the two-dimensional (2D) case in which the gravitational acceleration, magnetic field and direction of wave propagation are all co-planar (in the $x$--$z$ plane for example). In $(x,y,z)$ Cartesian coordinates, we set the background magnetic field to $\B_0=B_0(\sin\theta,0,\cos\theta)$, where $\theta$ is the inclination angle of the field relative to the vertical (see Figure \ref{fig:WidthsAndHeights}).

Before describing this exact solution in the remainder of this section, we pause to 
make a few contextual remarks. By exact solutions for MAG waves we mean solutions
where the analogous $\Psi$s are given in terms of tabulated special functions. There 
are many such exact solutions in the literature. Virtually all of these pertain to an
equilibrium where the magnetic field is everywhere \emph{ perpendicular} to the gravitational
acceleration and the atmospheric stratification. Such a configuration admittedly 
constitutes a set of measure zero when compared with all the possible directions a
magnetic field may point. Moreover, this case is also singular in the sense that neither the
mathematical methods nor physical outcomes follow uniformly from the $\theta \to
\pi/2$ limit of the \citeauthor{ZhuDzh84aa} exact solution, where the governing differential equation reduces from fourth order to second order, with attendant perplexing introduction of horizontal \emph{critical layers}. In these
layers wave energy-fluxes may be discontinuous (a process unfortunately called
`resonant absorption') and some components of the wave motion are unbounded, or diverge
as one approaches the layer from above or below. Steep gradients develop. 
The divergence of a wave amplitude is,
of course, entirely at odds with the linearization procedure. Some authors invoke 
finite dissipation to limit spatial gradients, suppress divergences, and provide a physical
basis for the resonant absorption. Such a `renormalization' of an infinity is not
unreasonable. Yet, it leaves several fundamental questions unanswered. The horizontal field case is discussed briefly in Section \ref{sec:horiz}.

The  $\theta \rightarrow
\pi/2$ limit of Zhugzhda and Dzhalilov's exact solution is particularly valuable in
providing the correct {\em physical} interpretation of the singular 
{\em mathematical} behavior of these horizontal magnetic field MAG wave solutions. This
limit also answers the useful question as to just how inclined a magnetic field needs
to be in order that it is effectively horizontal.

For all these reasons we devote this section to the remarkable exact solution of \citeauthor{ZhuDzh84aa} despite, as shall presently become clear, the significant amount
of algebra and special-function gymnastics involved. 

In terms of the component of plasma velocity $\uperp$ perpendicular to the background magnetic field
(lying in the $x$--$z$ plane), the linearized MHD equations may be combined into a single fourth-order ODE best couched in terms of dimensionless versions of parameters familiar from the above acoustic-gravity solution. 

First, introduce the independent position variable $s=\omega H/a$ in terms of the wave circular frequency $\omega$, density scale height $H$ and Alfv\'en speed $a(z)=a_0 \exp[z/2H]$; the dimensionless frequency $\nu=\omega H/c$; the dimensionless {\bv} frequency $n=NH/c$; and the dimensionless horizontal wavenumber $\kappa=k_xH$. Note that $s$ increases \emph{downward}, from $s=0$ at $z=+\infty$ to $s=\infty$ at $z=-\infty$. We then introduce the dimensionless vertical wavenumber from the acoustic-gravity dispersion relation Equation~(\ref{acGravDisp}),
\begin{equation}\label{kappa_z}
    \kappa_z=\sqrt{\nu^2-\kappa^2+n^2\kappa^2/\nu^2-\quart},
\end{equation}
where it should be noted that the $\quart$ is the square of the acoustic cutoff frequency in these units. Finally, we define $\kappa_0=\sqrt{\nu^2\sec^2\theta-\frac{1}{4}}$, the significance of which will become clear shortly.

The perturbed variables in the linearized MHD equations may all be eliminated in favour of $\uperp$, resulting in a linear homogeneous  fourth-order ODE \citep[see][for both a statement of the DE and solutions in terms of Meijer G-functions]{ZhuDzh84aa}
\begin{multline}
\uperp^{(4)} s ^4+4 \sec \theta  (\cos \theta -i \kappa  \sin \theta ) \uperp^{(3)} s ^3\\
+2
   \sec ^2\theta  \left[\left(2 \kappa_0^2+1\right) \cos ^2\theta -4 i \kappa  \sin \theta  \cos\theta
   +2 \left(s ^2-\kappa ^2\right)\right] \uperp'' s ^2\\
   +4 \sec ^2\theta  \left[4 \,i \, \kappa ^3\cos \theta 
   \sin \theta -\kappa ^2+3 s ^2+\cos ^2\theta  \left(2 \kappa ^2+\kappa_0^2\right)\right]
   \uperp' s\\
    +4 \sec ^2\theta  \left[4 \kappa ^4-4\, i \cos \theta  \sin \theta\,  \kappa ^3-\cos
   ^2\theta  \left(4 \kappa ^2+4 \kappa_0^2+1\right) \kappa ^2+s ^2 \left(4 \kappa _z^2+1\right)\right]
   \uperp\\
   =0.            \label{exactDE}     
\end{multline}
As in our previous cases, we have once again omitted writing out the explicit 
expression of the source term that should appear on the right side of this equation. 

This equation, while daunting in appearance, is actually amenable to an analytic
treatment. One begins by noting that this equation has two singular points: a regular
singularity at the origin ($s=0$) and an irregular singularity at infinity ($1/s=0$).
In a neighborhood of the former, the method of Frobenius may be applied to determine
four linearly-independent power-series solutions. Alternatively, the fourth-order differential equation can be written in a standard form known to admit hypergeometric solutions.

\citet{Cal01aa} for vertical magnetic field and \citet{HanCal09aa} for the general case recognized that these four solutions may be expressed most simply in terms of the $\Fh$ generalized hypergeometric function,
\begin{equation}
\Fh(a_1,\, a_2;\, b_1,\, b_2,\, b_3;\, x) =
\sum_{n=0}^\infty
\frac{(a_1)_n \, (a_2)_n}{(b_1)_n \, (b_2)_n \, (b_3)_n}\,
\frac{x^n}{n!} \, ,                                        \label{2F3}
\end{equation}
where $(a)_0=1$, $(a)_n=a(a+1)\ldots(a+n-1)$ is the Pochhammer symbol. These functions are entire. Specifically, they found
\begin{equation}
\begin{split}
\uperp = & C_1\, u_1+C_2\,u_2+C_3\,u_3+C_4\,u_4\\[6pt]
= & C_1\, s^{-2\kappa}\, 
\Fh\bigl(\ts\half - \kappa - i \kpz,\, \half - \kappa + i \kpz;
\\   & \qquad\quad
 \, 1-2\kappa,\, \half - \kappa - \ri\kappa_0-\ri\kappa\tan\theta,\, \half - \kappa + \ri\kappa_0-\ri\kappa\tan\theta;
\, -s^2\sec^2\theta\bigr) 
\\[6pt]
&+ C_2\, s^{2\kappa} \, 
\Fh\bigl(\ts\half + \kappa - i \kpz,\, \half + \kappa + i \kpz;
\\   & \qquad\quad
\, 1+2\kappa,\, \half + \kappa - \ri\kappa_0-i\kappa\tan\theta,\, \half + \kappa + \ri\kappa_0-\ri\kappa\tan\theta;
\, -s^2\sec^2\theta\bigr) 
\\[6pt]
&+ C_3\, s^{1-2\ri\kappa_0+2\ri\kappa\tan\theta} \, 
\Fh\bigl(\ts1-\ri\kappa_0-i \kpz+\ri\kappa\tan\theta,\,1-\ri\kappa_0+i \kpz+\ri\kappa\tan\theta;
\\   & \qquad\quad
\,1-2i\kappa_0,\,\threeontwo-\ri\kappa_0-\kappa+\ri\kappa\tan\theta,\,\threeontwo-\ri\kappa_0+\kappa+\ri\kappa\tan\theta;
\,-s^2\sec^2\theta\bigr) 
\\[6pt]
&+ C_4\, s^{1+2\ri\kappa_0+2\ri\kappa\tan\theta} \, 
\Fh\bigl(\ts1+\ri\kappa_0-i \kpz+i\kappa\tan\theta,\,1+\ri\kappa_0+i \kpz+\ri\kappa\tan\theta;
\\ & \qquad\quad
\,1+2\ri\kappa_0,\,\threeontwo+\ri\kappa_0-\kappa+\ri\kappa\tan\theta,\,\threeontwo+\ri\kappa_0+\kappa+\ri\kappa\tan\theta;
\,-s^2\sec^2\theta\bigr)\, ,
                                                           \label{uperp}
\end{split}
\end{equation}
where the $C_i$ are four arbitrary amplitudes. As the $\Fh$ power series all
tend to 1 as $s\to0$, the behavior of each solution for small $s$ is
determined by the power of $s$ multiplying the hyper\-geometric function in each expression. 

The solutions $u_1$ and 
$u_2$ are paired. They describe a single MAG wave mode with opposite directions of 
propagation (toward or away from $s=0$). Likewise, $u_3$ and $u_4$ pair to describe
a second distinct MAG wave mode. These two wave modes are asymptotic to the fast
and slow MHD waves, respectively, in the WKBJ limit. The direction of propagation
is determined by the sign convention chosen for the Fourier transforms in $t$ and
$x$. In other words, $\uperp$ is multiplied by a factor $\exp[\pm i(k_x x \pm
\omega t)]$ in calculating the inverse Fourier transform with 
a definite choice for each $\pm$. We adopt $\exp[i(k_x x -
\omega t)]$ throughout. Typically, one member of
each pair is dropped because it leads to an acausal or unphysical solution in
the neighborhood of $s=0$.

The family of generalized hypergeometric functions, $\Fz(\{a_i\},\{b_j\};x)$ includes
most of the familiar special functions of mathematical physics, for example,
the exponential ($p=q=0$), Bessel ($p=0, q=1$), Whittaker ($p=q=1$), and Legendre
($p=2, q=1$), functions. When $p$ is less than or equal to $q$, as obtains above, then
the power series is absolutely convergent for all finite values of s. (Otherwise,
$x=1$ is a third regular singularity, and the power series converge only for $|x|$ 
less than one.) What is especially valuable about this family of functions is that
each power series may be expressed as a contour integral. This in turn may be used 
to determine the asymptotic behavior of the power series in the neighborhood of
the irregular singular point $1/s=0$. In other words, one obtains the asymptotic
behavior of a given power series as a linear combination of the four 
linearly-independent solutions valid in the neighborhood of the irregular singular
point at infinity! 

Analytic expressions for the four coefficients -- expressed in
terms of the coefficients $\{a_i\}$ and $\{b_j\}$ -- are conveniently provided by \citet{Luk75aa}. 
Let $U_j$ for $j = 1, 2, 3, 4$ be the four linearly-independent solutions in the
neighborhood of $1/s = 0$, then for each $i = 1, 2, 3, 4$ we have $u_i=a_{ji}U_j$ with summation convention implied, i.e., 
\begin{equation}
    u_i = a_{1i}\, U_1 + a_{2i}\, U_2 + a_{3i}\, U_3 + a_{4i}\, U_4,
    \label{connection}
\end{equation}
where the sixteen $a_{ji}$ are known in terms of $\kappa$, $\kappa_0$, $\kappa_z$ and $\theta$ and are independent of $s$. All sixteen are set out in Table 1 of \citet{HanCalDon16aa}, and involve nothing more complicated than gamma and trigonometric functions.

Because $1/s=0$ is an irregular singular point of the ODE, the expressions for
the $U_j$ are formally divergent infinite series which are asymptotically exact
as $s\rightarrow \infty$. This behavior is entirely analogous to the asymptotic
expressions for the Hankel functions $H_\nu^{(1)}(x)$ and $H_\nu^{(2)}(x)$ as
$x \rightarrow \infty$, i.e., 
\begin{equation}
    H_\nu^{(1,2)}(x) \sim \sqrt{\frac{2}{\pi x}} \exp\left[\pm\ri
    \left(x-\frac{\nu\pi}{2}
    -\frac{\pi}{4}\right)\right].
\end{equation}

In the present circumstances each of the four hyper\-geometric solutions with Frobenius solutions centred at $s=0$ connects to \emph{four} asymptotic behaviours \citep[Equation (16.11.8)]{NIST:DLMF}
\begin{multline}
u_i \sim 
a_{1i}\,s^{-1/2+2 i \kappa\tan\theta}\, e^{2i s\sec\theta}+
 a_{2i}\,s^{-1/2+2 i \kappa\tan\theta}\, e^{-2i s\sec\theta}
  \\
+ a_{3i} \, s^{-1+2 i \kpz} + a_{4i} \, s^{-1-2 i \kpz}
                          \qquad  \mbox{ as $s\to\infty$.}                     \label{jasym}
\end{multline}
As before, these four asymptotic leading orders form fast (1 and 2) and slow (3 and 4) mode pairs, downward and upward propagating respectively for each pair.  

With these sixteen connection coefficients $a_{ji}$ in hand, 
the path forward is clear. Suppose, for example, 
one has specified some source term and needs to
determine the Green's function for a delta-function source at the point
$s=s'$. For $s \le s'$ we express the Green's function as a linear combination
of the two causal/physical solutions valid in the neighborhood of $s=0$.
For $s \ge s'$ we take a different linear combination of the two causal/physical
solutions with asymptotic expansions valid in the neighborhood of $1/s=0$. This 
leaves us with four undetermined complex constants. We use the four linear
equations in Equation (\ref{connection}) to express our two causal/physical 
asymptotic solutions as a definite linear combination of the $u_i$. This
is possible because all the $a_{ji}$ are known. Now the Green's function 
on both sides of the source $s=s'$ is expressed in terms of the same set
of four $u_i$. To determine the four unknown coefficients, we require that
the Green's function and its first three derivatives (with respect to s) be continuous at
$s=s'$, and that the fourth derivative is discontinuous by the appropriate
amount to generate the delta-function. 
This uniquely determines the Green's function, which can now be inverse-Fourier-transformed
along with the source term.

The essential point here is that a WKBJ fast (or slow) mode with a certain 
direction of propagation at large (or small) $s$ is invariably coupled to both
WKBJ modes with both propagation directions at small (or large) $s$. In a certain
sense, the input of one of the four solutions at one end of the atmosphere results
in the output of all four solutions at the other end. This behavior is conveniently
referred to as `mode-conversion' or `mode-coupling'.

As pointed out by \citet{GooArrVan19aa}, MHD waves in non-uniform plasmas exhibit `mixed properties', in that their physical characteristics -- fast or slow, acoustic or magnetic -- may differ in different regions of the atmosphere, even for a single global solution. This is alternatively interpreted as `mode conversion' by \citet{SchCal06aa} and \citet{CalGoo08aa}, but the idea is the same. The exact hypergeometric solutions presented above afford us an ideal test bed for understanding these phenomena. For example, the general but approximate fast/slow mode conversion theory of \citet{SchCal06aa} (see Section \ref{sec:fast/slow}) was compared to the exact hypergeometric solutions in an isothermal atmosphere by \citet{HanCal09aa}, and found to perform very well.

As shown below, the coupling between fast and slow waves occurs near the equipartition level where sound and Alfv\'en speeds coincide, $a=c$. In dimensionless units, this corresponds to $s=\nu$. The important question, therefore, is how the waves in $s\gg\nu$ and $s\ll\nu$ are coupled. For the isothermal atmosphere, this can be found by examining the asymptotic behaviours of the solutions in those regimes.

The Green's function is constructed to permit only outgoing (causal) MAG waves at 
either end of the atmosphere. To get at the global MAG modes without solving for the
Green's function, or specifying a source term, we ask the following question. Is it
possible to find a solution of the homogeneous equation where the amplitude of one
incoming mode is prescribed at one end of the atmosphere and only outgoing MAG
waves are permitted at each end of the atmosphere?

The small $s$ (large positive $z$) regime is easily accessed: all the $\Fh$ functions in the general solution tend to 1, as seen from Equation (\ref{2F3}). Hence,
\begin{equation}
\uperp\sim C_1\, s^{-2\kappa} + C_2\, s^{2\kappa} + C_3\, s^{1-2i \kappa_0+2i \kappa\tan\theta} + C_4\, s^{1+2i \kappa_0+2i \kappa\tan\theta},\mbox{ as $s\to0$.}
\end{equation}
The physical nature of the four constituent solutions is made clear on returning to $z$ as the independent variable: they are respectively the exponentially-growing evanescent fast mode, the decaying evanescent fast mode, the outgoing slow (acoustic) mode, and the incoming slow mode. So, for example, if a wave is being driven from below with no input from above, we must set $C_1=C_4=0$. 

The significance of $\kappa_0=\left(\nu^2\sec^2\theta-{1/4}\right)^{1/2}$ is now apparent. In the low-$\beta$ regime $a\gg c$ it becomes imaginary for $\nu<\half|\cos\theta|$, or in dimensional units $\omega<\omega_c|\cos\theta|$, indicating that the slow wave is evanescent below this inclination-modified acoustic cutoff frequency (the \emph{ramp effect}). This was first noted by \citet{BelLer77aa}, and invoked by \citet{JefMcIArm06aa} as opening `magneto\-acoustic portals' at supergranule boundaries for low frequency waves below $\omega_c$ to heat the solar chromosphere.

\begin{figure}[htb]
    \centering
    \includegraphics[width=1.0\textwidth]{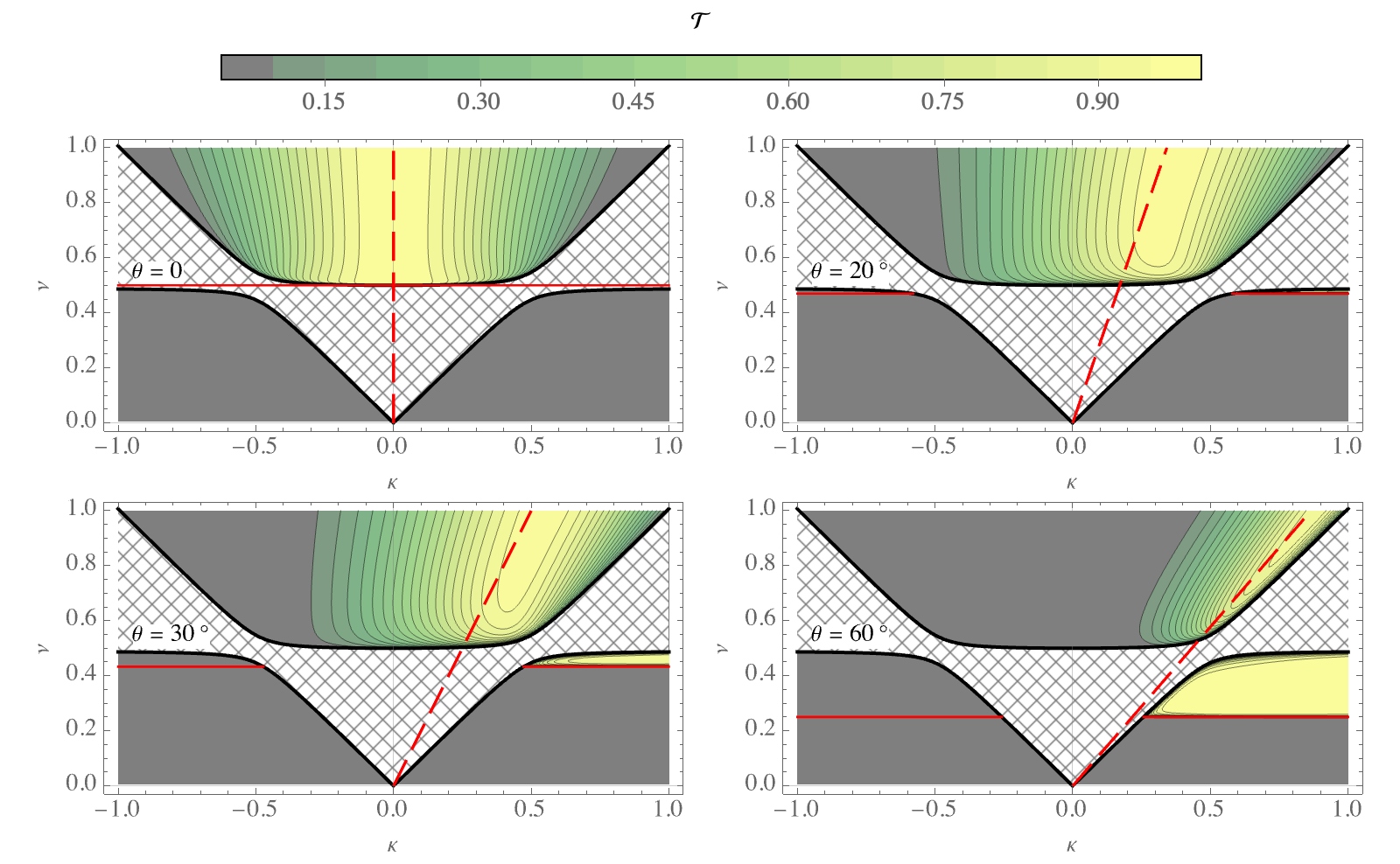}
    \caption{Exact global transmission coefficient $\mathcal{T}$ as a function of dimensionless wave number $\kappa$ and frequency $\nu$ for a 2D fast wave injected  from below into an isothermal atmosphere with uniform inclined magnetic field, for four different magnetic field inclinations $\theta$: top left $0^\circ$; top right $20^\circ$; bottom left $30^\circ$; bottom right $60^\circ$. No such injection is possible in evanescent Regions III and IV (hatched). The transmitted wave reaching the top is purely slow (acoustic) as the fast wave is evanescent there. The horizontal red line indicates the ramp-reduced acoustic cutoff frequency $\half\cos\theta$ below which the top acoustic wave cannot propagate. The dashed oblique red line is $\nu=\kappa \csc\theta$, corresponding to the direction of maximum transmission predicted by ray theory (Section~\ref{sec:fast/slow}). Figure constructed using the EMAGWIA package of \citet{HanCalDon16aa}.}
    \label{fig:T}
\end{figure}

To summarize, we first express the solution in terms of four hyper\-geometric functions $\uperp=C_i u_i$ (summation convention implied). These functions are defined in terms of power series centred at $s=0$ ($z=+\infty$), and have easily interpretable physical natures there. Alternatively, we could just as well use as the basis the four pure asymptotic behaviours as $s\to\infty$ ($z=-\infty$), $\uperp=c_i U_i$. Thanks to our exact solutions, the alternate basis coefficients $\bc=(c_1,\ldots,c_4)$ and $\C=(C_1,\ldots,C_4)$ are connected via $\bc=A\C$ where the non-singular matrix $A=(a_{ij})$ is known in terms of gamma functions. The $4\times4$ matrix $A$ therefore provides full coupling information between the two asymptotic regimes, allowing easy calculation of reflection transmission and conversion coefficients \citet{HanCalDon16aa}.

For example, Figure~\ref{fig:T} shows the fractional transmission coefficient $\mathcal{T}\in[0,1]$ (transmitted top flux divided by injected bottom flux) over-plotted on the acoustic-gravity propagation diagram Figure~\ref{fig:propdiag} for an injected fast (acoustic or gravity) wave from below and a range of magnetic field inclinations $\theta$. This transmission is purely in the form of acoustic (Region I) or gravity (Region II) waves, as the fast (i.e., magnetic) wave is evanescent at the top. It is global in the sense that it takes account of the cumulative effects of reflection in $a<c$, transmission through $a=c$, and reflection in $a>c$. Both positive and negative $\kappa=k_xH$ are included to address waves propagating with and against the direction of magnetic field inclination. Points to note include: (i) the maximal transmission is roughly along the line $\nu=\kappa \csc\theta$, which corresponds to minimal attack angle between wave vector and magnetic field at $a=c$ (see Section~\ref{sec:fast/slow}); (ii) there is little transmission `against the grain' of significantly inclined field, i.e., with $\kappa<0$; and (iii) low frequency gravity waves (Region II) with $\kappa>0$ and frequency $\nu>\half\cos\theta$ transmit almost totally due to the ramp effect, though there is little flux for $\kappa<0$ because there is little gravity-wave-to-gravity-wave transmission at $a=c$ for that case. Below the ramp frequency, the gravity waves reflect in the $a\gg c$ strong-field region (see Equation (\ref{ramp})).

Although this model is very idealized, these features are characteristic of fast and slow waves in more general strongly stratified atmospheres. Of course, we have yet to discuss Alfv\'en waves.

\subsection{Decoupled Alfv\'en Waves}\label{sec:exactAlf2D}
Alfv\'en waves are understood to play an important role in heating the solar atmosphere and heating and accelerating the solar wind \citep{Cravan05aa}.\footnote{As well as Alfv\'en wave propagation, their dissipation is also an important part of the story. Being incompressive, they are not strongly affected directly by most diffusive mechanisms of solar relevance. However, plasma turbulence and hence heating may be generated via nonlinear counter-propagating interactions if there is a source of Alfv\'en reflection in the atmosphere \citep{MatZanOug99aa,HowNie13aa}. This will not be discussed further here.} Yet, to feed the corona and solar wind they must penetrate the roughly six orders of magnitude photosphere/chromosphere density contrast, not to mention the precipitously steep gradients of the transition region (see Figure~\ref{fig:TnH}). What can exact solutions tell us about their ability to do so?

In a uniform plasma, the fast, slow and Alfv\'en waves are all independent of each other and have mutually orthogonal velocity polarizations. Vertical stratification couples the fast and slow magneto-acoustic waves as seen in the 2D exact solutions of Section \ref{sec:exact2D}, where the direction of wave propagation and the magnetic field are in the same vertical plane, say $x$--$z$. However, the Alfv\'en wave is polarized perpendicular to this plane, and is entirely decoupled from the other two wave types. Because the Alfv\'en polarization is horizontal, the plasma displacement does not interact with the stratification and there are no buoyancy forces operating on it, so there is no compression. This means that the Alfv\'en wave is incompressive, $\Div\boldv=0$.

With this in mind, and noting that both $\boldv$ and $\B_1$ are polarized in the $y$-direction, the linear momentum equation reduces to
\begin{equation}
    \rho_0\pderiv{v_y}{t}=\frac{1}{\mu}\,\B_0\,\vdot\,\grad\,B_{1y}
\label{Alf mmntm}
\end{equation}
and the induction equation to
\begin{equation}
    \pderiv{B_{1y}}{t}=\B_0\,\vdot\,\grad\, v_y . 
\end{equation}
Combining the two yields
\begin{equation}\label{AlfDEdd}
    \rho_0\pderivd{{v_y}}{t}=
    \frac{1}{\mu}
    (\B_0\,\vdot\,\grad)^2v_y ,
\end{equation}
for any $\B_0(x,z)$.
If $\B_0$ is uniform, then this reduces to the classic 1D wave equation,
\begin{equation}\label{Alf eqn}
    \pderivd{v_y}{t}=(\a\,\vdot\,\grad)^2v_y=a^2\pderivd{v_y}{\sigma}
    =a^2\cos^2\theta\,\pderivd{v_y}{z}\biggr|_\text{f.l.},
\end{equation}
where $\a=a\,\hat\B_0$ is the Alfv\'en velocity, $\sigma$ is distance along the field line, $d\sigma=\sec\theta\,dz$, the $z$-derivatives are along a field line (f.l.), and $\theta$ is the magnetic field inclination from the vertical.

For the same isothermal atmosphere discussed in Section \ref{sec:exact2D}, where $a^2=a(0)^2 \exp(z/H)$ with $a(0)$ the Alfv\'en speed at $z=0$, and again assuming an $\exp i(k_xx-\omega\,t)$ dependence, Equation (\ref{Alf eqn}) has exact solutions in terms of Bessel functions of order zero, $J_0$ and $Y_0$, or Hankel functions $H_0^{(1)}=J_0+i\,Y_0$ and $H_0^{(2)}=J_0-i\,Y_0$:
\begin{equation}\label{AlfHank}
    v_y = s^{2i\kappa\tan\theta}\left(V_1 \, H_0^{(1)}(2s\sec\theta) + V_2 \, H_0^{(2)}(2s\sec\theta)\right),
\end{equation}
where $\kappa=k_xH$ and $s=\omega H/a$ is the same dimensionless spatial coordinate as used there, and $V_1$ and $V_2$ are arbitrary complex constants with dimensions of velocities. The two Hankel functions represent respectively waves travelling in the positive and negative $s$-directions, in other words downward and upward. The $J_0=\half(H_0^{(1)}+H_0^{(2)})$ solution alone is a standing wave, a superposition of two equal-amplitude oppositely directed Hankel solutions.

The period-averaged wave energy flux carried by an Alfv\'en wave is simply the Poynting flux $\mu^{-1}\E_1\vcross\B_1$, where $\E_1=-\boldv\vcross\B_0$ is the electric field perturbation, the vertical component of which gives 
\begin{equation}\label{Fz}
  \begin{split}
    f_z &=\frac{B_0}{\mu}\re\left[B_{1y}v_y^*\right]\cos\theta \\
    &= \frac{B_0^2}{\mu\,\pi\,H\,\omega}\left(|V_2|^2-|V_1|^2\right)\cos^2\theta
  \end{split}  
\end{equation}
in the complex representation.
Note that $f_z$ is independent of $z$, as it must be if energy is conserved.

It is of interest to determine how this flux is affected by the scale-height $H$. Does a smaller scale-height, and hence a steeper Alfv\'en speed gradient, reduce the flux of wave energy that the Alfv\'en wave can carry? Let us assume that a purely upgoing Alfv\'en wave is driven at $z=0$ with a given energy density $\mathcal{E}=B_1^2/2\mu$. Then $|V_2|^2 = 2\mu B_0^{-2}\mathcal{E}\, a(0)^2|H_1^{(2)}(2\omega H a(0)^{-1}\sec\theta)|^{-2}$, and hence
\begin{equation}\label{Fz driven}
    f_z = 
    a(0)\,\mathcal{E}\cos\theta\,\varepsilon_\text{A}(X),
\end{equation}
where $X=2s(0)\sec\theta=(2\omega H/a_0)\sec\theta$ and $\varepsilon_\text{A}(X)=2(\pi\,X)^{-1}|H_1^{(2)}(X)|^{-2}\sim1$  as $X\to\infty$. 

\begin{figure}[htb]
    \centering
    \includegraphics[width=0.65\textwidth]{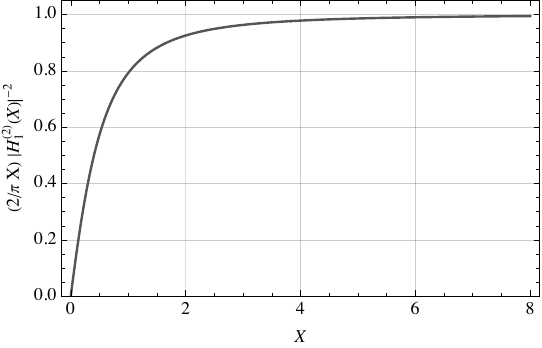}
    \caption{The Alfv\'en transport efficiency function $\varepsilon_\text{A}(X)=2(\pi\,X)^{-1}|H_1^{(2)}(X)|^{-2}$, where $X=(2\omega H/a_0)\sec\theta$.}
    \label{fig:fX}
\end{figure}

The Alfv\'en transport efficiency function $\varepsilon_\text{A}(X)$ is plotted in Figure~\ref{fig:fX}. It shows that the flux $f_z$ is constant, $f_z\sim  a(0)\cos\theta\,\mathcal{E}$ as $(\omega H/a(0))\sec\theta\to\infty$, which is just the uniform atmosphere value, and also the WKBJ value. This is the maximal efficiency that can be achieved: an energy density $\mathcal{E}$ transported vertically at projected Alfv\'en speed $a(0)\cos\theta$ is precisely this asymptotic value.

On the other hand, $f_z\to0$ as $X\to0$ (low frequency, or small scale height, or large base Alfv\'en speed, or large field inclination), so these `long dimensionless wavelength' cases are significantly less efficiently transported due to an Alfv\'en speed gradient that they perceive as steep.  The optimal transport speed is not attained as $X$ gets small and the solution becomes less wave-like: $H_0^{(2)}(X)\sim -(2i/\pi)\ln X$ as $X\to0$. This contrasts with the transparently wavelike $H_0^{(2)}(X)\sim\sqrt{2/\pi X}\exp[-i(X-\pi/4)]$ as $X\to\infty$. Note that $X$ increases with field inclination, via the term $\sec\theta$, and consequently efficiency increases, due to the wave experiencing a less steep Alfv\'en speed slope as it propagates along the field line.

To put this in context in the most magnetic environment available on the Sun's surface, a sunspot umbra, let us consider umbral Model M of \citet{MalAvrCar86aa}. At $z=0$ (optical depth unity) in the umbral core, the density scale height is about 145 km. With a magnetic field strength of 3 kG the Alfv\'en speed is 8.1 $\rm km\,s^{-1}$. Adopting a typical umbral wave period of 180 seconds then corresponds to $X=1.2$, which gives an efficiency $\varepsilon_\text{A}(X)\approx0.85$. By period 300~s, characteristic of the wider solar surface, this drops to 0.7, suggesting that even lower frequency Alfv\'en waves are difficult to excite in sunspot umbrae. This is not an issue in weaker magnetic field regions where $X\gg1$.

In reality of course, Alfv\'en speeds in the solar atmosphere do not increase exponentially with height indefinitely. Once in the corona, the density scale becomes much larger. It is therefore of interest to explore Alfv\'en speed profiles $a(z)$ that initially rise exponentially but then plateau to a uniform value \citep{Cal12aa}. This (abrupt or smooth) change of scale height produces some reflection, with coefficient $\mathcal{R}$ being the ratio of the reflected wave energy flux to the injected flux. Using WKBJ methods, it is found that $\mathcal{R}=\mathcal{O}\left(\omega^{-2d}\right)$, where $d$ is the order of the discontinuity of slope. For example, if $a'(z)$ changes discontinuously at some point, then $d=1$ and reflection decreases quadratically with increasing frequency. The smoother the profile and the higher the frequency, the smaller the reflection. 

For a $\mathscr{C}^\infty$ function (i.e., infinitely differentiable) the reflection is exponentially, not algebraically, small. For example, an exact solution is available for the analytic (and therefore infinitely differentiable) profile $a^2=a(0)^2 e^{z/h}/(1+\epsilon\,e^{z/h})$, for which the plateau is at $a_u=a(0)/\sqrt{\epsilon}$. This yields an exact exponentially small reflection coefficient $\mathcal{R}=\exp(-4\pi k_u h)$, where $k_u=\omega/a_u$ is the wave number in the uniform plateau.

These mathematical niceties may seem a long way from the real Sun, but they do nevertheless indicate that there can be reflection of Alfv\'en waves propagating from photosphere to corona to solar wind, with more reflection at low frequencies and for more abrupt Alfv\'en speed changes of slope. The determining parameters are the wavenumber, the Alfv\'en scale height and the degree of smoothness of the Alfv\'en speed profile. \citet{Vel93aa} postulates that multiple such reflections could create leaky Alfv\'en wave cavities, where they are imperfectly contained.

Initial-value solutions of Equation (\ref{Alf eqn}) for non-uniform Alfv\'en speed almost inevitably exhibit \emph{wakes} or \emph{reverberations} that typically decay as $\mathcal{O}(t^{-1})$ or $\mathcal{O}(t^{-1}\ln t)$ (depending on initial condition) associated with a continuous spectrum \citep{Cal12aa}. Wakes are partially reflective. Except for a few very specific profiles $a(z)\propto z^{(n-1)/(n-2)}$ with odd-integer $n$, reverberations are characteristic of Alfv\'en waves propagating in atmospheres of inhomogeneous Alfv\'en speed.

\subsection{Horizontal Magnetic Field, Critical Levels, Resonant Absorption and Phase Mixing}\label{sec:horiz}
Horizontal magnetic field is an obvious and important case to explore for waves in the solar atmosphere. Magnetic field emerging from small (e.g., network elements) and large (sunspots) flux concentrations at the photosphere rapidly spreads covering the quiet solar surface with highly inclined field. The level at which the plasma beta, $\beta=p/p_\text{mag}$, equals 1 is called the `canopy' and separates regions of different behaviour. Sunspot penumbrae may also be modelled as regions of (tolerably) horizontal field. This overlying blanket of magnetic field may also have small cycle-dependent helio\-seismic effects \citep{CamRob89aa}.

The study of waves in horizontal fields $\B_0=(B_0(z),0,0)$ affords us the opportunity to introduce some interesting concepts. A nice mathematical discussion of linear waves in a stratified atmosphere with horizontal field is given by \citet{NyeTho76aa}, who write the governing differential equation for the vertical velocity perturbation $w$ in the form
\begin{equation}\label{Nye DE}
    \derivd{\hat w}{z}+A(z)\,\deriv{\hat w}{z}+B(z)\,\hat w=0,
\end{equation}
having made the standard $\boldv=\hat\boldv(z)\exp i(k_xx+k_yy-\omega\,t)$ assumption for perturbation velocity $\boldv(\x,t)=(\hat u(z),\,\hat v(z),\,\hat w(z))$.

The most striking point to notice, compared to the inclined field case of Equation (\ref{exactDE}) is that the differential equation is only of second order, rather than fourth order. This is the nub of the distinction. Something has disappeared!

The coefficients $A$ and $B$ deepen the mystery:
\begin{multline}\label{Nye A}
    A(z)=-\frac{1}{H}+\frac{\omega^4}{D\,E}(\omega^2-a^2k_x^2)^2 \deriv{c^2}{z}\\[4pt]
    -\frac{1}{D}
    \left[\omega^4-(\omega^2-c^2k_x^2)(\omega^2-a^2k_x^2)\left(1+\frac{\omega^4}{E}\right)\right]\deriv{a^2}{z}
\end{multline}
and
\begin{multline}\label{Nye B}
    B(z)=\frac{1}{D}\Biggl\{ \omega^6-\left[(a^2+c^2)k^2+a^2k_x^2\right]\omega^4-a^2k_x^2k^2\left[a^2c^2k_x^2-g\left(g-\frac{c^2}{H}\right)\right]\\
    +\left[a^2k_x^2k^2(2c^2+a^2)-g k^2\left(g-\frac{c^2}{H}\right)+\frac{g}{H}a^2k_y^2 \right]\omega^2 \\[4pt]
    -\frac{g}{E}(\omega^2-a^2k_x^2)^2\omega^2k^2\deriv{c^2}{z}-\frac{\omega^6}{E}g k_y^2 \deriv{a^2}{z}\Biggr\}=0,
\end{multline}
where $k_x^2+k_y^2=k^2$, $g$ is the gravitational acceleration and $H$ is the density scale height. The most interesting terms though are 
\begin{equation}\label{Nye D}
    D=(a^2+c^2)(\omega^2-a^2k_x^2)(\omega^2 -\cT^2k_x^2)
\end{equation}
where $\cT=a\,c/\sqrt{a^2+c^2}$ is the cusp speed, and
\begin{equation}\label{Nye E}
    E=\omega^4-(a^2+c^2)k^2\omega^2+a^2c^2k^2k_x^2.
\end{equation}
The reader may recognize $E$ as the magneto-acoustic factor in the MHD dispersion function in a uniform plasma, Equation (\ref{MHD disp}).

\citeauthor{NyeTho76aa} obtain solutions in terms of ${}_2F_1$ hypergeometric functions for the special case of uniform sound speed and magnetic field with $k_y=0$, but we will leave them there. Of more interest to us are these terms $D$ and $E$ that appear in the denominator in $A(z)$ and $B(z)$. These two coefficients are singular wherever $D$ or $E$ vanish, indicating something strange is happening. Alternatively, we can multiply the whole Equation (\ref{Nye DE}) through by $DE$ to make the coefficients analytic, but the vanishing of the coefficient of the highest derivative suggests that the solutions are singular.

The singular points inherent in $DE$ are where the fast and slow dispersion function vanishes, $E=0$, and where the Alfv\'en and cusp relations $\omega^2=a^2k_x^2$ and $\omega^2=\cT^2k_x^2$ apply, $D=0$. Application of the Frobenius series expansion method reveals that the fast and slow singularities \emph{do not} produce singular $\hat w$ solutions, so are not of concern. However, solutions for $\hat w$ resulting from $D=0$ \emph{are} singular; each has both an analytic and a $\ln(z-z_0)$ singular solution in the neighbourhoods of their respective singular points, $z_\text{A}$ and $\zT$. The vertical wave flux $f_z\propto \re(\hat p_T\hat w^*)$, where $\hat p_T$ is the necessarily continuous total pressure perturbation, so the 
logarithm results in a discontinuity in wave flux, since $\ln(z-z_0)=\ln|z-z_0|+i\arg(z-z_0)$, which is interpreted as absorption. 

These two problematic critical points are called the Alfv\'en and cusp \emph{critical layers} or \emph{resonances}. They have appeared because of the vanishing of the coefficient of the second derivative, and the non-existence of higher derivatives seen in the non-horizontal field case.

Alfv\'en resonant absorption was discussed in the laboratory plasma context by \citet{CheHas74aa} and by \citet{Ion78aa} for coronal loops. The topic is reviewed in depth by \citet[][Chap.~11]{GoePoe04aa}, and in Chapter 5 of this volume. We will not go into further mathematical detail here. However, we take the opportunity to explain why the singular nature of the solutions is due to rather idealized assumptions, in particular (i) that the wave has been propagating forever, and (ii) that it is spread over all $x$. If any of these two assumptions is relaxed, the singularity disappears and the physical nature of `resonant absorption' becomes simpler.

\begin{figure}[htb]
    \centering
    \includegraphics[width=.75\textwidth]{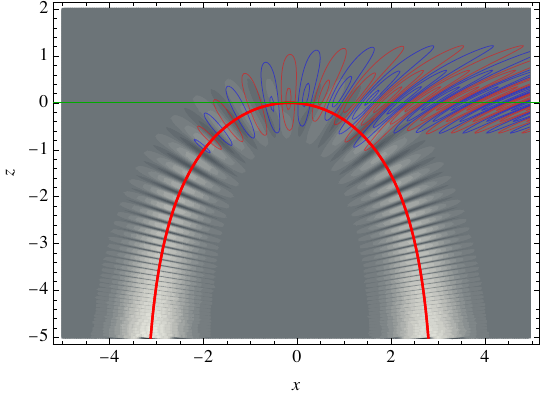}
    \caption{Dilatation $\chi=\Div\bxi$ for a fast wave Gaussian ray bundle (grey-scale) is launched from around $x=-\pi$ in a cold-plasma ($c=0$) stratified atmosphere with uniform horizontal magnetic field in which Alfv\'en speed increases exponentially with height. Here $\bxi$ is the plasma displacement vector. The red curve indicates the path of the central ray of the bundle, for which $k_x=9$ and $k_y=3$, and the green horizontal line is the Alfv\'en critical level for that central ray. By the process of `mode conversion' or `resonant absorption', as you prefer, about half the energy is converted from the fast wave to an Alfv\'en wave, with its displacement $\xi_y$ rendered in blue and red (towards and away from the viewer). This Alfv\'en bundle then propagates to the right along field lines. }
    \label{fig:chiRot}
\end{figure}

For example, consider a compact incident fast ray bundle \citep{CalAnd10aa}. In Figure \ref{fig:chiRot} a compact bundle of fast rays is launched upward in a stratified atmosphere with uniform horizontal magnetic field, reflecting around $z=0$. Each ray in the bundle has a slightly different launch position and horizontal wavenumber $k_x$. The Alfv\'en critical level of each ray is slightly above its turning point. The fast and Alfv\'en waves interact only if $k_y\ne0$, since otherwise there is no coupling mechanism. Wave energy is partially converted to Alfv\'en waves that then propagate away to the right along field lines, travelling at their own local Alfv\'en speeds at each height $z$. 

Hence, the upper portion of the bundle moves faster than the bottom portion, causing the Alfv\'enic transverse-velocity blue and red phase regions to progressively incline further as the wave progresses. This results in ever-larger velocity gradients, and is called \emph{phase mixing}. Ultimately, it leads to thermalization via non-ideal processes such as friction. The opportunity for the Alfv\'en waves to escape the conversion region prevents the build-up of infinite velocities seen in the horizontally-invariant case, and is also the reason that there are no singularities when the field is even slightly inclined, since in that case too the Alfv\'en waves can escape. A similar conclusion is drawn for the case of a horizontally invariant wave pulse of finite duration, which leaves only a finite amount of energy on the Alfv\'en resonances, and which phase mixes in time \citep{HanCal11aa}. 

Phase mixing was explored by \citet{HeyPri83aa} for a much simpler system that beautifully illustrates its nature. If you think of the strings of a harp, like taut magnetic field lines with different Alfv\'en speeds, and pull them all out to one side and let go, the higher frequency strings will vibrate faster and neighbouring strings, no matter how close, will eventually find themselves completely out of phase with each other. Hence the name `phase mixing' is very apt. This creates arbitrarily fine spatial structure, and hence the opportunity for dissipation and even Kelvin-Helmholtz instability \citep{TerAndGoo08aa,AntVan19aa}. 

The process is more complex when the plasma velocities are not perpendicular to the direction of Alfv\'en speed inhomogeneity, but the principle is the same, there is a cascade of energy to ever finer scales \citep{Cal91ab}. In that sense, despite being a linear process, phase mixing shares some characteristics with 3D turbulence.

\subsection{Exact 3D MHD Solutions} \label{sec:exact3D}
In three dimensions, where gravity, magnetic field and wave direction are not co-planar, all three MHD wave types -- fast, slow and Alfv\'en -- are coupled, at least in certain regions, so fast/slow, fast/Alfv\'en and even slow/Alfv\'en conversion are implicit.

The full three-dimensional (3D) sixth-order differential system for MHD waves in an isothermal atmosphere with uniform inclined magnetic field was written down as two coupled third order equations by \citet{ZhuDzh84aa}. Using the same dimensionless variables as in Section \ref{sec:exact2D}, complicated recurrence relations were found for Frobenius expansions about $s=0$. Leading order asymptotic expansions about $s=\infty$ were also sketched, but not continued to completion. \citeauthor{ZhuDzh84aa} noted that no closed-form exact solution was available.

\citet{CalGoo08aa} wrote the governing sixth order equation in a more elegant matrix form $s \U'=A\,\U$, where $A=A_0+s^2A_2$ and $\U=(\xi,\eta,\zeta, s\xi', s\eta', s\zeta')^T$, where $\bxi=(\xi,\eta,\zeta)$ is the displacement vector in Cartesian form. They used this to develop the full Frobenius solutions about $s=0$.

Recently, \citet{Cal22in}, using the same formulation, also constructed the complete asymptotic solution about $s=\infty$, making extensive use of matrix algebra. 

At this stage, there are no known exact coupling coefficients between the two asymptotic regimes,\footnote{The industrious reader may wish to take this as a challenge!} so no exact transmission and conversion coefficients. However, \citet{Cal22in} coupled them numerically using the Frobenius and asymptotic solutions to explore all three mode conversion types (between the three mode types) over ranges of frequencies, wave numbers and magnetic field inclinations. We do not present the results here, but emphasise that substantial mode conversion of all three types does indeed occur in 3D magneto-atmospheres (see Section~\ref{sec:overall}).

\section{Eikonal Method and Ray Theory}  \label{sec:eikonal}
The description of MHD waves in terms of wavevectors, frequencies and dispersion relations discussed in Section \ref{sec:waves} is applicable in basic form only to uniform media. For more complex systems, such as the stratified and magnetically structured atmospheres encountered in solar physics, numerical simulations are often the method of choice. Those based on finite differences, finite volumes or finite elements have no direct link to those Fourier-based methods. Spectral methods \citep{CanHusQua88aa} only use them in a global sense, with solutions constructed using a large collection of modes. The most natural successor is ray theory, where rays are paths in $\x$ space that carry solutions based on slowly evolving $\omega$, $\k$ and their local dispersion relations. This is constructed from the same eikonal foundation as the WKBJ method. In this section, we give a brief overview of ray theory and show how it applies to MAG waves. This will then prove useful conceptually and practically in describing mode conversion.

\subsection{Ray Theory Basics}\label{sec:ray basics}
Ray-based methods, also called geometrical optics, are applied widely in many disciplines. Based on the same eikonal underpinnings as their WKBJ cousin, they essentially model a wave as a moving particle, leveraging the familiar wave-particle duality concept from quantum mechanics.

The purpose of this approach is to reduce the formidable partial differential equations (PDE) of fluid dynamics or MHD to ordinary differential equations (ODE) along rays. These are much more easily and quickly integrated numerically, though come with their own attendant difficulties. Ray methods inherit their degree of accuracy or inaccuracy from the eikonal assumption, and so are most applicable when wavelengths are small compared to the scales over which the background atmosphere is changing. Ray methods also apply for backgrounds that slowly vary in time, assuming that wave periods are small with respect to background evolution times. An excellent discussion of the foundations and applications of ray theory in fluids may be found in \citet{Whi65aa}, and an entertaining array of applications is presented by \citet{Ada17aa}.

Within magneto\-hydro\-dynamics, the fundamental ideas are set out by \citet{Wei62aa}, with \citet{TraBriRic14aa} focusing more widely on plasma physics and on modern developments. The starting point may be either the governing PDEs (\citeauthor{Wei62aa}) or a variational formulation (\citeauthor{TraBriRic14aa}). The former is more familiar to most solar physicists, though the latter introduces powerful insights and tools. 

We proceed with the PDE approach. On linearizing the continuity, momentum, induction and energy equations, a system of linear PDEs in space $\x$ and time $t$ results for variables such as velocity $\boldv_1$, magnetic field perturbation $\B_1$, density and pressure perturbations $\rho_1$ and $p_1$, etc. These are placed on top of the slowly varying (in space and possibly time) background quantities $\boldv_0(\x,t)$, $\B_0(\x,t)$, $p_0(\x,t)$, etc. 

The nub of the method is to assume that the perturbation quantities, representing the waves, can be expressed in terms of a slowly varying amplitude and a common rapidly varying phase $S(\x,t)$, e.g.,
\begin{equation}\label{eik expand}
    \B_1(\x,t) = \B_{10}(\x,t)\,e^{i\,S(\x,t)}, \quad \rho_1(\x,t)=\rho_{10}\,e^{i\,S(\x,t)},
\end{equation}
etc. Space and time derivatives of the amplitudes (subscripts `10') are ignored in comparison to those of $S$, but their $\x$ and $t$ dependencies are retained nonetheless in the coefficients of the resulting differential equations.

With that in mind, a wave vector $\k$ and angular frequency $\omega$ are defined by
\begin{equation}\label{komega def}
    \k=\grad S \quad\text{ and }\quad \omega=-\pderiv{S}{t}.
\end{equation}
Formally, since only derivatives applying to $S$ are retained, this identifies $\grad\equiv i\,\k$ and $\partial/\partial t\equiv -i\,\omega$ when applied to perturbation variables, and reduces calculus to algebra. The resultant system of simultaneous linear algebraic equations can then be written in an $n\times n$ matrix form
\begin{equation}\label{M ray}
    M(\x,t;\k,\omega)\,\bpsi=\mathbf{0}
\end{equation}
for the column vector $\bpsi$ of retained dependent variables, such as the components of $\boldv_{10}$, $\B_{10}$, etc. Some of these variables may be eliminated in favour of others if convenient, which reduces the size of the matrix and vector. The $\x$ and $t$ retained in $M$ are the slowly varying dependencies of the background atmosphere.

Equation (\ref{M ray}) has non-trivial solutions only if 
\begin{equation}\label{D ray}
    \mathcal{D}(\x,t;\k,\omega) = \det\bigl(M(\x,t;\k,\omega)\bigr) = 0.
\end{equation}
This is the dispersion relation, familiar from Section \ref{sec:waves}, but now with slow $\x$ and $t$ dependence. The null vectors $\e$ of $M$ are the slowly evolving polarizations of the eigensolutions, and only have meaning on the $\mathcal{D}=0$ manifolds.

Suppose we are on a ray, which is a curve in physical and phase space $(\x,t;\k,\omega)$ parameterized by some parameter $\tau$ of no particular physical relevance. The ray lies on the manifold $\mathcal{D}=0$. Many different rays lie on $\mathcal{D}=0$, associated with differing starting positions. Assume our particular ray is parameterized as $\left(\x(\tau),t(\tau);\k(\tau),\omega(\tau)\right)$. Differentiating the dispersion relation implicitly with respect to $\tau$ gives
\begin{equation}\label{D difft}
    \pderiv{\mathcal{D}}{\x}\,\vdot\,\deriv{\x}{\tau}
    +\pderiv{\mathcal{D}}{t}\,\deriv{t}{\tau}
    +\pderiv{\mathcal{D}}{\k}\,\vdot\,\deriv{\k}{\tau}
    +\pderiv{\mathcal{D}}{\omega}\,\deriv{\omega}{\tau} = 0,
\end{equation}
where the total time derivative $d/d\tau$ is the rate of change moving along the ray. 

Equation (\ref{D difft}) is true for any motion on the dispersion manifold, so it only partially restricts the ray path. Let us further impose that the direction of $d\x/d\tau$ should be in the direction of the group velocity, $\partial\mathcal{D}/\partial\k$. An arbitrary scaling determines $\tau$. Similarly, $\k$ should change in the direction $\grad\mathcal{D}=\partial\mathcal{D}/\partial\x$, since $\k$ is invariant if $\mathcal{D}$ is independent of $\x$. This is sufficient to constrain the characteristic ray equations:
\begin{subequations}\label{ray eqns}
\begin{align}
       &\deriv{\x}{t}=-\frac{\partial\mathcal{D}/\partial\k}{\partial\mathcal{D}/\partial\omega},\label{rayx}\\[4pt]
   & \deriv{\k}{t}=\frac{\partial\mathcal{D}/\partial\x}{\partial\mathcal{D}/\partial\omega},\\[4pt]
   & \deriv{\omega}{t} = -\frac{\partial\mathcal{D}/\partial t}
       {\partial\mathcal{D}/\partial\omega}.
\end{align}
\end{subequations}

Rapid phase evolution may be integrated in parallel with the ray equations (\ref{ray eqns}),
\begin{equation}\label{S evoln}
   \deriv{S}{t}=\grad{S}\,\vdot\,\deriv{\x}{t} + \pderiv{S}{t} = \k\,\vdot\,\deriv{\x}{t}-\omega,
\end{equation}
 having applied Equations (\ref{komega def}). In cases where $\mathcal{D}$ is independent of $t$ and so $\omega$ is constant, it is common practice to assume an overall $e^{-i\omega t}$ dependence and leave off the $-\omega$ term in $dS/dt$. 

It remains to evaluate the slowly varying vector amplitude 
\begin{equation}\label{psi}
    \bpsi(\x,t)=A(\x,t)\,e^{i\phi(\x,t)}\,\e(\x,t),
\end{equation}
where $A>0$ is a scalar amplitude and $\phi$ is a slowly varying real phase correction. The unit polarization vector $\e$ is simply the normalized null vector of $M$, which varies (slowly) with position along the ray.

Determining the amplitude relies on the action conservation law
\begin{equation}\label{action cons}
    \pderiv{\mathscr{I}}{t}+\div\!\left(\mathscr{I} \boldv_\text{gr}\right)=0,
\end{equation}
where $\mathscr{I}=\mathscr{E}/\omega$ is the wave action, $\mathscr{E}=A^2\omega\,\partial D_{\!\alpha}/\partial\omega$ is the wave energy density and $\boldv_\text{gr}$ is the group velocity, which is simply the right hand side of Equation (\ref{rayx}). Here $D_{\!\alpha}$ is the particular eigenvalue of $M$ that belongs to the ray in question, and may be taken as the dispersion function for just that ray. Alternatively, the full dispersion function $\mathcal{D}$ may be used instead. In the case where the background atmosphere is steady and frequency is constant along rays, $\omega$ may be cancelled, leaving an energy conservation equation. However, action conservation remains true even if $\omega$ is slowly varying, and in that sense Equation (\ref{action cons}) is more general and more fundamental \citep[Section~11.7]{Whi65aa}. The conservation law may be rewritten in the form
\begin{equation}\label{action cons d}
   \deriv{\ln\mathscr{I}}{t}+\Div\boldv_\text{gr}=0,
\end{equation}
where the total time derivative $d/dt=\partial/\partial t+\boldv_\text{gr}\vdot\,\grad$ follows the ray. This illustrates the point made previously about slowing and focusing of rays increasing action, and therefore amplitude. Integrating Equation (\ref{action cons d}) in parallel with Equations (\ref{ray eqns}) determines $A$ along the ray, given an initial value at the start point. 

Finally, the equation for determining the slow phase evolution requires a higher order asymptotic analysis too involved to go into here \citep[see Section~3.4 of][for details]{TraBriRic14aa}:
\begin{equation}\label{d phi}
    \deriv{\phi}{t} = i\, \e^*\vdot\, \deriv{\e}{t} +\half i\, M_{lm}\left\{e_l,e_m^*\right\},
\end{equation}
where $M_{lm}$ are the components of $M$, $e_l$ are the components of $\e$, the summation convention is implied, and $\{f,g\}=\grad_\x\, f\,\vdot\,\grad_\k\, g-\grad_\x \,g\,\vdot\,\grad_\k\, f$ is the Poisson bracket. 

To summarize, given a starting point $\x_0$, and starting frequency $\omega$ and wave vector $\k$ consistent with $\mathcal{D}=0$, Equations (\ref{ray eqns}) for $\x$, $\k$ and $\omega$; Equation (\ref{S evoln}) for $S$; Equation (\ref{action cons d}) for $A$; and Equation (\ref{d phi}) for $\phi$, may be integrated numerically in time to determine its path, amplitude and phase. This is repeated for an ensemble of rays, starting from an initial launch surface with phase that is consistent with the eikonal ansatz. For example, if launching from $z=0$, one might set $S(0)=k_x x+k_yy$ and $\phi(0)=0$. In this way, a full solution my be constructed by interpolation between finely spaced rays.

However, there are two major issues with these solutions.
\begin{enumerate}
    \item \textbf{Caustics:} Rays in $\x$-space are projections from the full $\x$--$\k$ (and possibly $\omega$) space in which they more properly exist. This means that rays may cross in 3D physical space despite being separated in the full 6D phase space, leading to an unphysical infinite amplitude. Although both $\x(t)$ and $\k(t)$ remain valid through these \emph{caustics}, both phase\footnote{\citet{Bog97aa} presents an interesting comparison of the wave and ray descriptions in helioseismology, and in particular discusses the phase jump at turning point caustics.} and amplitude are incorrect and require special treatment at problematic junctures along the ray integrations \citep[Ch.~5]{TraBriRic14aa}. The phase jump depends solely on the topological nature of the projection singularity, but anyway, this is all very inconvenient. See \citet[Appendix 16]{Arn89aa} for mathematical details of caustics. Very recently, a new technique called Metaplectic Geometrical Optics \citep[MGO,][]{LopDod22aa}, which works in mixed $\X=A\x+B\k$ coordinate space for particular evolving matrices $A$ and $B$, has been developed in which the caustics do not appear at all, and so the ray equations may be integrated without interruption. Treatment of caustics is far too complex for us to go into here, but the modeller needs to be aware of the issue, which arises in solar applications due to refraction and reflection.
    \item \textbf{Mode Conversion:} The MHD dispersion relation describes fast, slow and Alfv\'en waves as distinct modes. However, we saw in Section \ref{sec:exact2D} that waves in stratified atmospheres exhibit mixed properties, otherwise known as mode conversion. The transmissions/conversions happen when ray paths pass close to each other in $\x$--$\k$ space, enabling resonant interactions (see Section \ref{sec:conv}). Basic ray theory does not describe this phenomenon. However, at least for fast/slow interaction \citep{SchCal06aa}, a generalization of it does (see Section \ref{sec:fast/slow}).
\end{enumerate}

With these caveats in mind, let us look at MHD rays in particular.

\subsection{Application to MHD Waves}\label{sec:ray apps}
\subsubsection{Dispersion Relation}\label{sec:dispRel}

It is useful to retain the acoustic cutoff and buoyancy in any low frequency ray integration, notwithstanding any loss of accuracy due to $k H$ not being large. To do this, it is common practice to include both $\omega_c$ and $N$ is the dispersion function. Depending on variables used, the precise magneto-acoustic-gravity dispersion relation can take various forms, even including imaginary terms \citep{McLWin68aa} despite being adiabatic. The imaginary components describe variations in amplitude, which we access via wave action conservation instead.

A real MAG dispersion relation should reduce to the classic MHD result (\ref{MHD disp}) for uniform magnetic field in the absence of gravity, and to the acoustic-gravity relation Equation (\ref{acGravDisp}) in the absence of magnetic field. Neglecting derivatives of $\B$, a suitable form is derived by \citet{NewCal10aa}:
\begin{multline}\label{MAG disp 2D}
    \left(\omega^2-a^2\kpar^2\right)
    \left(\omega^4 - (a^2+c^2)\omega^2 k^2+a^2c^2k^2\kpar^2+c^2N^2k_\text{h}^2-(\omega^2-a_z^2k^2)\omega_c^2\right)\\
    +\omega^2\omega_c^2 a_\text{t}^2k_h^2
    =0,
\end{multline}
where $k_h$ is the horizontal wave number, $a_z=B_z/\sqrt{\mu\rho}$ is the $z$-projected Alfv\'en speed, and $a_\text{t}$ is the component of the Alfv\'en velocity $\a$ transverse to both $\k$ and $\mathbf{g}$. Hence $k_h^2a_\text{t}^2=|\k\,\vcross\,\e_z\,\vdot\,\a|^2$. 

This dispersion relation has the desirable property that it recovers the ramp effect for field-guided acoustic waves in the $a\gg c$ limit, 
\begin{equation}\label{ramp}
    \omega^2 = c^2\kpar^2+\omega_c^2\cos^2\theta,
\end{equation}
where $\theta$ is the angle of the magnetic field from the vertical. That is, the effective cutoff frequency is reduced by the factor $\cos\theta$ \citep{BelLer77aa}.\footnote{Beware the typographical error in \citet{BelLer77aa}, Eq.~(8), where $\cos\theta$ should be squared.}

\begin{figure}[htb]
    \centering
    \includegraphics[width=0.92\textwidth]{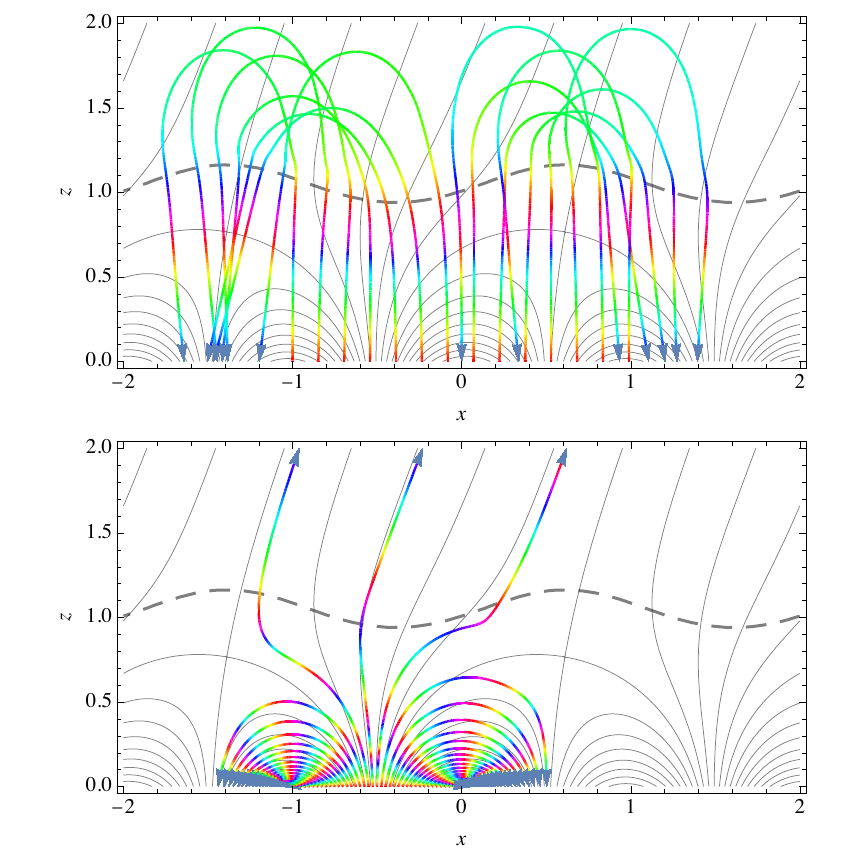}
    \caption{A 2D isothermal potential-field magneto-atmosphere with $c=8\,\rm km\,s^{-1}$, $g=274\,\rm m\,s^{-2}$, $H=c^2/\gamma g=140.15$ km and $\gamma=5/3$. Magnetic field is specified by Equation (\ref{AOC}) with $B_\text{u}=32$~G and $B_0=355$~G, period $L=2$ Mm and $\theta=20^\circ$. The full black curves are magnetic field lines and the dashed curve is the equipartition surface, $a=c$. Horizontal $x$ and vertical $z$ distances are in Mm. Top: a selection of fast 6 mHz rays launched from $z=0$ with $k_x=0$. Bottom: a selection of slow 6 mHz rays launched from $z=0$ with $k_x=0$. Arrows show the directions of the rays. Colours indicate phase.}
    \label{fig:rays}
\end{figure}

\subsubsection{MAG Rays in Open and Closed Field}\label{sec:OC}
Consider a two-dimensional (2D) isothermal equilibrium atmosphere with magnetic field $\B_0=(\partial A/\partial z,0,-\partial A/\partial x)$ of $x$-period $L=2\pi/K$ where 
\begin{equation}\label{AOC}
    A(x,z)=B_\text{u} (z\sin\theta-x\cos\theta) -  \frac{B_0}{K} \, e^{-K z}\cos K x.
\end{equation}
This represents a mixture of open and closed magnetic field.

For simplicity, we launch fast and slow rays with horizontal wave numbers $k_x=k_y=0$ and frequency $\omega$ from $y=0$, $z=0$ at various $x$ positions. Since the background is independent of $y$ and $t$, neither $k_y$ or $\omega$ change along the rays. As $a_\text{n}=0$ the magneto-acoustic and Alfv\'en waves decouple and the dispersion function may be taken as
\begin{equation}\label{D rays}
   \mathcal{D}=\omega^4 - (a^2+c^2)\omega^2 k^2+a^2c^2k^2\kpar^2+c^2N^2k_x^2-(\omega^2-a_z^2k^2)\omega_c^2.
\end{equation}
The atmospheric parameters are $c=8\,\rm km\,s^{-1}$, $g=274\,\rm m\,s^{-2}$, $H=c^2/\gamma g=140.15$ km and $\gamma=5/3$. The acoustic cutoff and {\bv} frequencies are respectively 4.54 mHz and 4.45 mHz. For the magnetic field we set $B_\text{u}=32$ G, $B_0=355$ G, $L=2$ Mm and $\theta=20^\circ$.

Figure \ref{fig:rays} depicts fast and slow 6 mHz rays launched with $k_x=0$ from various positions on the base. The fast rays all reflect above $a=c$ around where their horizontal phase speed matches the local Alfv\'en speed, and do not differ significantly depending on whether they are in open or closed magnetic field. This is due to the basic property of fast waves that they travel freely in all directions (see Figure \ref{fig:fried_ph} and \ref{fig:fried_gr}). Fast waves are predominantly acoustic in $a\ll c$ and predominantly magnetic where $a\gg c$.

On the other hand, slow waves are largely bound to the magnetic field (Figure \ref{fig:fried_gr}), and so are trapped in closed field but escape in open field. Slow waves are predominantly magnetic in $a\ll c$ and predominantly acoustic, albeit field-guided, where $a\gg c$.

These ray calculations were carried out using Equations (\ref{ray eqns}) and (\ref{S evoln}) only. The phase correction $\phi$ was not applied. The amplitude $A(x,z)$ was not evaluated. Caustics were not accounted for. No account was taken of mode conversion at $a=c$; see \citet{SchCal06aa} for examples of integrations that incorporate ray splitting. We turn to mode conversion in Section \ref{sec:conv}.

Despite these limitations, the solutions depicted, which altogether took well under a second to calculate on a laptop, present useful information on the behaviour of MAG waves in a highly stratified and magnetically structured atmosphere. Compared to large finite difference, finite elements or finite volumes codes, they are much easier and faster to run, are cheap to extend to 3D, are not subject to spurious reflections at the boundaries, and are trivial to parallelize, as each ray is independent. Ray solutions can also be easier to interpret as the different wave types are kept separate, though of course this can be a double-edged sword.


\section{Mode Conversion and Coupling} \label{sec:conv}

Mode conversion came to prominence in the context of solar p-mode absorption
\citep{BraDuvLab87aa,BraDuvLab88aa}, though initially the mechanism was uncertain. \cite{Spr91aa} and \cite{SprBog92aa} were the first to suggest that the incident f- and p-modes were being converted to slow MHD waves that disappeared down field lines into the solar interior. This was subsequently confirmed by theoretical modelling \citep{CalBogZwe94aa}, simulation \citep{Cal00aa} and detailed comparison with scattering data 
\citep{Bra95aa,CalCroBra03aa}.

Subsequently, a new generation of powerful simulation codes also started producing mode conversion between fast and slow modes in chromospheric models \citep{BogCarHan03aa}, apparently associated with the `magnetic canopy' which was generally understood to be where the plasma-beta ($\beta=p/p_\text{mag}$) was unity, or almost equivalently, where the sound and Alfv\'en speeds coincided. This was later verified and understood using a generalised ray theory by \cite{Cal06aa} and \cite{SchCal06aa}.

\cite{CalGoo08aa} further explored fast/slow conversion numerically, but also discovered substantial fast/Alfv\'en conversion in their models, provided they were 3D, i.e., gravity, magnetic field and wave direction were not co-planar. This provided an important new route for waves to potentially reach and heat the corona, since fast waves typically reflect from the steep Alfv\'en speed gradient in the high chromosphere and cannot themselves penetrate higher in a plane-stratified atmosphere. Alfv\'en waves, produced by mode conversion in the chromosphere, do not suffer this problem.

The story of fast wave reflection and conversion was further enriched by their suggested and ultimately verified role in explaining seismic halos around active regions \citep{KhoCol09aa,RijRajPrz16aa}.

\begin{figure}[htb]
    \centering
    \includegraphics[width=0.6\textwidth]{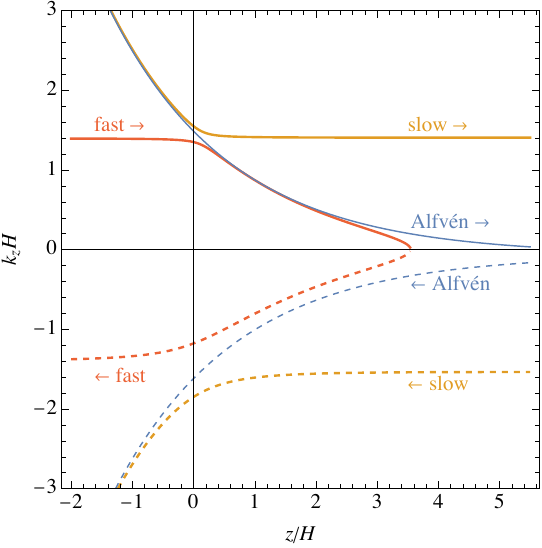}
    \caption{Dispersion diagram for isothermal atmosphere with uniform magnetic field inclined $\theta=15^\circ$ from the vertical and oriented $\phi=10^\circ$ out of the $x$--$y$ plane. The dimensionless frequency $\nu=\omega H/c$ is 1.5 (compared to the acoustic cutoff frequency of 0.5) and $k_x H=0.25$ where $H$ is the density scale height.The fast, Alfv\'en and slow loci are labelled and coloured individually, with dashed lines corresponding to downward propagating waves. The $a=c$ equipartition level is $z=0$.}
    \label{fig:DispDiag}
\end{figure}

In this section we briefly explore the natures of fast/slow and fast/Alfv\'en conversion. This is aided by $z$--$k_z$ dispersion diagrams like Figure \ref{fig:DispDiag}. 

With the magnetic field inclined in the same general direction as the wave propagation (positive $x$ in this case), there is a close avoided crossing between the fast and slow waves on the upgoing legs ($k_z>0$) near the equipartition level $a=c$ (arbitrarily identified with $z=0$ here). This is where strong fast/slow mode \emph{transmission} occurs, i.e., the waves are enabled to jump the gap from fast to slow (acoustic to acoustic) or from slow to fast (magnetic to magnetic). The somewhat wider gap on the downgoing legs allows weaker transmission.

On the other hand, the fast and Alfv\'en dispersion loci are almost coincident over a long distance on the upgoing leg above $z=0$. This would seem to give ample opportunity for fast-to-Alfv\'en (or vice-versa) resonant conversion, provided the magnetic field is oriented out of the $x$--$z$ propagation plane ($\phi\ne0$). However, it will be seen in Section \ref{sec:fast/Alf} that in fact the dominant interaction is in the evanescent tail above the fast wave turning point.

Again, the downgoing leg sees weaker interaction with this magnetic field orientation. If $\theta$ is changed to $-\theta$ (or equivalently $\phi$ to $\phi+\pi$) the picture is flipped, with interaction stronger on the downgoing legs.

These two mode conversion types are now discussed in more detail.

\subsection{Fast/Slow Conversion} \label{sec:fast/slow}

Employing the ray-based method of \citet{TraKauBri03aa}, \citet{SchCal06aa} derive an asymptotic transmission coefficient (acoustic-to-acoustic, i.e., fast-to-slow, or magnetic-to-magnetic, i.e., slow-to-fast) at avoided crossings of the type seen in seen in Figure \ref{fig:DispDiag}:
\begin{equation}\label{T}
    T = \exp\left[
    -\pi\, k \left|h_s\right| \frac{\sin^2\alpha}{1+\sin^2\alpha}
    \right]_*,
\end{equation}
where $\alpha$ is the attack angle between the wavevector and the magnetic field lines at a suitably selected `star point' in the gap of the avoided crossing, and $k=|\k|$ is the wave number.\footnote{Acoustic cutoff and {\bv} frequencies have been neglected in this derivation.} The length $h_s$ is the `thickness of the conversion layer' as encountered by the ray propagating at angle $\psi$ from the vertical, specifically $h_s=(d\ln(a^2/c^2)/dz)^{-1}\sec\psi$. Depending on exact formulation, the star point is typically at or near $a=c$, $k_z=\omega/c$ in phase space. Different choices of star point result in slightly altered formulae for $T$, but they all agree for small $\alpha$, where the gap is narrow. The corresponding conversion coefficient is $C=1-T$. 

It is important to point out that transmission is total for attack angle $\alpha=0$, but that at high frequencies, and hence large $k$, the `wedge of transmission' becomes progressively narrower in $\alpha$. That is, in a bath of variously directed waves, only those in a very narrow wedge of directions will transmit. For all others, conversion is almost total. That means that they will stay on their fast or slow branches, converting from acoustic to magnetic or \emph{vice versa} as they go, and not jump the gap. This is to be expected, because the eikonal description that sorts MHD waves into types (fast, slow or Alfv\'en) is more and more accurate as frequency increases. 

This formula for $T$ was tested against the exact solutions of Section \ref{sec:exact2D} and found to perform very well. It is also consistent with the numerical survey of wave transmission carried out by \citet{CalGoo08aa}.

\subsection{Fast/Alfv\'en Conversion} \label{sec:fast/Alf}
Fast-to-Alfv\'en conversion is not of the avoided crossing type, and therefore not amenable to the ray method used above. However, several other analytic and numerical approaches have mapped it out.

Perhaps the most instructive is the perturbation analysis of \citet{CalHan11aa}, which we adapt. Noting that the fast/Alfv\'en interaction typically occurs several scale heights above the equipartition layer $a=c$, it is reasonable to assume $c\ll a$, in which case the plasma displacement becomes entirely transverse, $\B_0\,\vdot\,\bxi=0$, and the governing equations reduce to
\begin{subequations}\label{parxi}
\begin{align}
\left(\partial_\parallel^2+\partial_\perp^2+\frac{\omega^2}{a^2}\right)\xi_\perp
 &= -i\, k_y \partial_\perp\xi_y \label{parxi f}\\
\left(\partial_\parallel^2+\frac{\omega^2}{a^2}\right)\xi_y &= -i\, k_y \partial_\perp\xi_\perp + k_y^2\xi_y \label{parxi A}.                                 
\end{align}
\end{subequations}
Here we assume the magnetic field lies in the $x$--$z$ plane, and the subscript `$\perp$' indicates the direction $(\cos\theta,0,-\sin\theta)$ in that plane and perpendicular to $\B_0$.

In the 2D case $k_y=0$, the two equations decouple and represent respectively pure fast and Alfv\'en waves. Hereafter we assume $k_y$ is small and neglect the $k_y^2$ term. Consider an uncoupled standing (i.e., reflecting) fast wave $\xi_{\perp0}(z)$ satisfying Equation (\ref{parxi f}) with zero right hand side (as in Figure \ref{fig:DispDiag}). Then we seek the Alfv\'en solution $\xi_y$ of inhomogeneous Equation (\ref{parxi A}) driven by $-i k_y\partial_\perp\xi_{\perp0}$. This is a first order regular perturbation approach. 

Consider an exponential Alfv\'en speed profile such that $a(z)=\omega H e^{z/2H}$ and introduce dimensionless wave numbers $\kappa=k_xH$ and $\kappa_y=k_yH$. Rescale $z$ so that $H=1$. The fast wave turning point $\omega=a k_x$ is therefore at $z_\text{tp}=-2\ln\kappa$.

The standing fast wave in terms of Bessel functions is
\begin{equation}
\xi_{\perp0}(z) = J_{2\kappa}(2\,e^{-z/2}) 
= \half\left(
H_{2\kappa}^{(1)}(2\,e^{-z/2})  + H_{2\kappa}^{(2)}(2\,e^{-z/2})
\right) =\xi_{\perp0}^- + \xi_{\perp0}^+  .\label{xiperp0}
\end{equation}
The arguments of the Bessel functions are $2\kappa$ at $z_\text{tp}$.

Using a Green's function, the first order perturbation solution for the Alfv\'en wave generated by $\xi_{\perp0}$ is
\begin{multline}
\xi_{y1} \sim  -\frac{\kappa_y\pi}{2}\,e^{-i\kappa z\tan\theta}\sec^2\theta\\
\times\left(
A(z)\,H_0^{(1)}(2e^{-z/2}\sec\theta) + B(z)\,H_0^{(2)}(2e^{-z/2}\sec\theta)
\right),
\end{multline}
where
\begin{equation}
\begin{aligned}
A(z) &= \int_z^\infty e^{i\kappa X\tan\theta}H_0^{(2)}(2e^{-X/2}\sec\theta)\,\partial_\perp\xi_{\perp0}(X)\,dX\, ,\\[4pt]
B(z) &= \int_{-\infty}^z e^{i\kappa X\tan\theta}H_0^{(1)}(2e^{-X/2}\sec\theta)\,\partial_\perp\xi_{\perp0}(X)\,dX
\end{aligned}  \label{AB}
\end{equation}
are the downgoing and upgoing \emph{interaction integrals}. It is assumed that there are no incoming Alfv\'en waves from above or below. 

\begin{figure}[htb]
    \centering
    \includegraphics[width=0.49\textwidth]{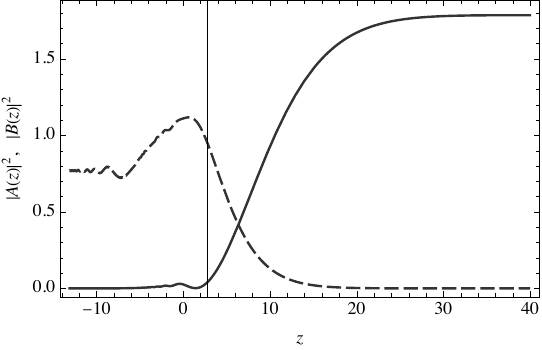}\hfill
    \includegraphics[width=0.49\textwidth]{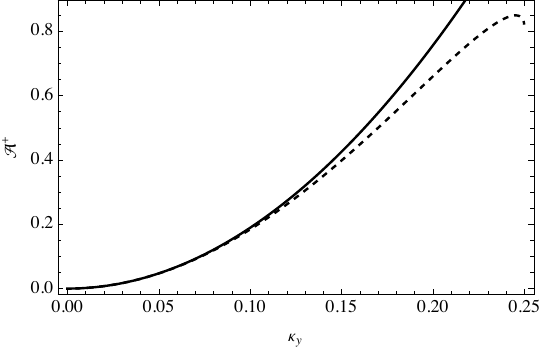}
    \caption{Left: absolute squares of the downgoing interaction integral $A(z)$ (dashed) and upgoing interaction integral $B(z)$ (full curve) for the case $\kappa=0.25$, $\theta=15^\circ$. This corresponds roughly to the case of Figure~\ref{fig:DispDiag}, notwithstanding the zeroing of sound speed here. The vertical line indicates the fast wave reflection height. Right: total upward Alfv\'en flux $\mathscr{A}^+\le1$ generated by resonant interaction with the injected fast wave, normalized by the injected flux. Full line: first order perturbation result; dashed line: exact flux.}
    \label{fig:ABinteract}
\end{figure}

The interaction integrals $A$ and $B$ are the amplitudes of the downgoing and the upgoing Alfv\'en waves generated by the resonance. The flux in each is proportional to $|A(z)|^2$ and $|B(z)|^2$ respectively. These are plotted in Figure \ref{fig:ABinteract}, left panel, for the `with the grain' case $\kappa=0.25$, $\theta=15^\circ$. This shows that $|B|^2$ increases from zero to a maximum plateau as $z$ increases over an extended range, indicating where resonance coupling is operating. Conversely, the downgoing power $|A|^2$ increases from zero at large $z$ in the tail to a lower plateau on negative $z$, again showing the interaction range, but also that the flux of downgoing Alfv\'en waves generated in this process is much smaller than the upgoing flux. This is expected since the fast wave has positive $k_x$ and $\theta$ is positive, which gives a closer coincidence of the fast and Alfv\'en loci on the upgoing leg. 

Notice also that the interaction starts just before but persists well above the fast wave reflection height ($z_\text{tp}=-2\ln\kappa=2.8$ in this case). This is because the evanescent tail of the fast wave is its only non-oscillatory part. Below $z_\text{tp}$, the rapid oscillations of the integrands of $A(z)$ and $B(z)$ result in almost total cancellation. This is especially so in the $A$ integral (assuming $\kappa\sin\theta>0)$, since the rapidly varying arguments of the fast and Alfv\'en waves add rather than subtract as they do in $B$.  Hence, the evanescent tail of the fast wave plays the dominant role, as it did in the `resonant absorption' horizontal field case of Figure \ref{fig:chiRot}.

The right panel of Figure \ref{fig:ABinteract} compares the total normalized Alfv\'en flux $\mathscr{A}^+\in[0,1]$ of the perturbation calculation with the exact value as a function of $\kappa_y$, showing that it performs well for sufficiently small $\kappa_y$. 

In summary, fast/Alfv\'en conversion is potent and concentrated at and above the fast wave reflection height. It is far more dispersed in height than fast/slow conversion.

\subsection{Net Mode Conversion in a Plane Stratified Atmosphere}\label{sec:overall}
It is useful to step back from an analysis of the individual processes discussed above, to gain an overview of how they combine to determine the acoustic and Alfv\'en fluxes reaching the upper atmosphere, depending on magnetic field direction, frequency and horizontal wavenumber.

\begin{figure}[htb]
    \centering
    \includegraphics[width=\textwidth]{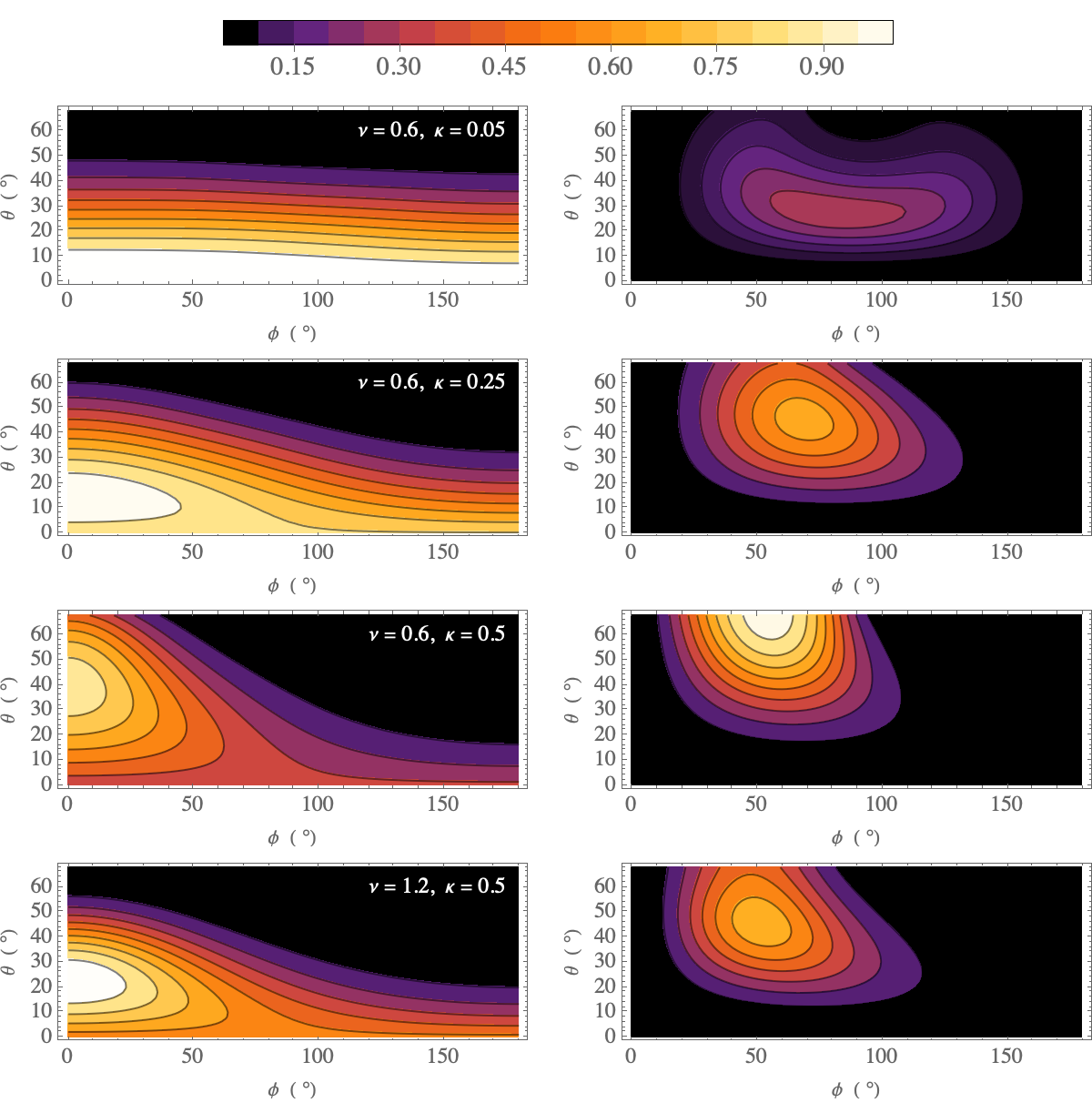}
    \caption{Acoustic (left) and Alfv\'en (right) fluxes normalized by the injection fast flux using the 3D exact solutions of Section \ref{sec:exact3D} with: $\nu=0.6$, $\kappa=0.05$ (top row); $\nu=0.6$, $\kappa=0.25$ (second row); $\nu=0.6$, $\kappa=0.05$ (third row); and $\nu=0.9$, $\kappa=0.25$ (bottom row). The cutoff frequency is $\nu_c=0.5$.}
    \label{fig:3DFlxs}
\end{figure}

Figure \ref{fig:3DFlxs} presents a survey against magnetic field inclination $\theta$ and orientation $\phi$ out of the $x$--$y$ plane of the acoustic and Alfv\'en fluxes, normalized by injected fast wave flux at the bottom, reaching the top in an isothermal atmosphere with uniform magnetic field, based on the exact solutions of Section \ref{sec:exact3D}. Recall the dimensionless frequency $\nu=\omega H/c$ and horizontal wavenumber $\kappa=k_xH$. The figure presents a distinctive picture of maximal acoustic flux at $\phi=0$, but weakening and moving to higher $\theta$ as $\kappa$ increases (first three rows). Increasing $\kappa$ corresponds to decreasing fast wave reflection height.

On the other hand, the Alfv\'en flux increases toward high inclinations and larger values of $\kappa$. In the last row, both $\nu$ and $\kappa$ are doubled compared to the second row, resulting in some translation of features to smaller $\phi$ but very little change in peak fluxes. This weak dependence on $\nu/\kappa=\omega/ck_x$, which is approximately the ratio of the Alfv\'en speed to the sound speed at the fast wave reflection height, was also seen in the 2D fluxes of Figure \ref{fig:T}.

These results are consistent with the numerical surveys of an active region related variable-temperature atmosphere by \citet{CalGoo08aa}. 

All panels in Figure \ref{fig:3DFlxs} relate to Region I of the acoustic-gravity propagation diagram Figure \ref{fig:propdiag}, for which $\nu>\max(\kappa,0.5)$. The top acoustic and magnetic fluxes do not sum to 1 as reflection (as fast, slow and Alfv\'en) also accounts for some of the injected energy: $f_\text{inj}=f_{\uparrow \text{A}}+f_{\uparrow \text{s}}-f_{\downarrow \text{A}}-f_{\downarrow \text{s}}-f_{\downarrow \text{f}}$, where the up-arrows denote outgoing fluxes at the top and the down-arrows indicate downgoing fluxes at the bottom. 

Perhaps the most important lesson to be drawn from these results is that Alfv\'en wave generation by acoustic-to-magnetic-to-Alfv\'en mode conversion is strongest in substantially inclined magnetic field ($\theta\approx30^\circ$-- $60^\circ$), and for waves crossing the vertical magnetic plane at high angle ($\phi\approx60^\circ$-- $80^\circ$).

\subsection{Wave Conversion around Null Points}
Till now, the $a=c$ equipartition/mode conversion surfaces we have encountered have been more-or-less horizontal. However, the complex magnetic structuring of the solar atmosphere opens new possibilities. An important example is the case of null points -- points where the magnetic field strength $B$ drops to zero. They are particularly prominent in 2D models, and take two forms: X-type, where field lines cross as seen in Figure \ref{fig:rays}, and O-type, where the field lines are essentially circular about a null centre. The topology of 3D null points is more complex \citep{ParSmiNeu96aa}, involving such structures as spines and fans, but for our purposes the main characteristic is that there is a point (or line) where the Alfv\'en speed drops to zero, and therefore, if it is embedded in an otherwise low-beta plasma such are the upper chromosphere or corona, there is a compact $a=c$ surface surrounding it where fast/slow mode conversion can, and indeed must, occur.\footnote{The fast rays are not drawn to the null point in Fig.~\ref{fig:rays} because it is placed well below $a=c$ and hence is not a zero or deep minimum of wave speed.}

An extensive review of MHD waves (fast, slow and Alfv\'en) around null points in 2D, 2.5D and 3D is presented by \cite{McLHoode-11lb}, and the reader is referred there for details. From our perspective, the important insight is that fast waves refract towards regions of lower Alfv\'en speed, and hence are attracted to and subsequently `wrap themselves around' null points embedded in low-beta plasmas \citep{McLHoo04aa} where they encounter $a=c$ and therefore conversion, or more properly a mix of conversion and transmission \citep{McLHoo06ab}. This is verified in the simulations of \citet{PenCal21dv}, and carries over to nonlinear and shock waves. Nonlinear waves of course disrupt and shred the neutral point, which makes neutral points and their surrounds prime sites for wave heating.
In this fashion, magnetic nulls are to fast MAG waves as black holes are to photons. 

\section{Outlook and Further Reading}
This chapter mainly focuses on theory and simple models in order to present general principles. Solar wave studies must relate to observations though. An up-to-date encyclopaedic review of waves in the solar photosphere and chromosphere from an observational and data analysis perspective is presented by \citet{JesJafKey23aa}, which is a good starting point for beginning researchers.

For reasons which should by now be fairly obvious, sophisticated numerical MHD codes are increasingly the
method of choice for modelling wave propagation in 2D and 3D magneto-atmospheres. Dozens of computationally-oriented 
papers are published each year reporting on the results of a diverse array of numerical simulations. 
The inquisitive reader is urged to sample this growing body of work. Owing to the
wide range of astrophysical applications, physical assumptions, and the variety of boundary and initial conditions
employed, no systemic assessment of their interrelationships and common findings is presently available.

The principal advantages offered by numerical simulations are (i)~the relative ease with which nonlinearities and
variations in space and time can be handled; (ii)~their capacity to be employed repetitively as an `experimental' tool; (iii)~the ability to add further physics such as radiation;
and (iv)~the insights they provide to answer the question we posed at the start of this article: 
\emph{What is a wave, and what is simply nonlinear dynamics}? On the other hand, computational methods are subject
to the vicissitudes of boundary conditions and resolution. These have posed, and will continue to pose, serious 
challenges, including, for example, spurious unphysical effects, and profound discord between results from different codes even for acoustic-gravity waves \citep{FleCarKho21aa}.
Accordingly, the methods and ideas presented here will always
be an essential weapon in the arsenal of the computational astrophysicist who wishes to test and interpret their results.

We close by mentioning a few additional extremely useful monographs devoted exclusively to waves in fluids 
that are subject
to some combinations of compressibility, buoyancy, rotation, and magnetic pressure/tension as their restoring forces.
Magnetized equilibria are addressed by \cite{Lif89aa}, \cite{Lic94aa}, \cite{Cra01aa}, \cite{04Walker},
\cite{07Alperovich}, \cite{Rob19aa}, while \cite{Bee74aa}, \cite{Lig78aa}, \cite{Cra88aa}, \cite{Ped03aa}, 
\cite{OckOck04aa}
treat buoyancy, compressibility, steady flows, and rotation.

\appendix
\renewcommand{\thesection}{\Alph{section}}
\section{Appendix: Thermodynamics}\label{app:thermo}

Following \citet{Eck60aa}, we record here some useful thermodynamic relations that enable one to choose a convenient form for the fluid internal energy equation. 

Let $\epsilon$ be the internal energy per unit mass (units: Joules kg$^{-1}$), 
$\upsilon$ (units: kg m$^{-3}$) the specific volume ($=1/\rho$), $p$ the pressure (units:
Joules m$^{-3}$), $T$ the temperature (units: Kelvins), and $\eta$ the
specific entropy (units: Joules kg$^{-1}$ Kelvin$^{-1}$). Then
\begin{equation}
    p = - \left( \pderiv{\epsilon}{\upsilon} \right)_\eta \qquad
    T = \left( \pderiv{\epsilon}{\eta} \right)_\upsilon,
\end{equation}
and the specific heats at constant volume and pressure are
\begin{equation}
    C_\upsilon = T \left( \pderiv{\eta}{T} \right)_\upsilon \qquad
    C_p = T \left( \pderiv{\eta}{T} \right)_p.
\end{equation}
The coefficient of thermal expansion and (the square of) the adiabatic
sound speed are
\begin{equation}
    \beta = \frac{1}{\upsilon} \left( \pderiv{\upsilon}{T} \right)_p \qquad
    c^2 =  \left( \pderiv{p}{\rho} \right)_\eta.
\end{equation}
Finally, the second partial-derivatives of $\epsilon$ are:
\begin{equation}
    \pderivd{\epsilon}{\upsilon} = \rho^2 c^2, \qquad
    \mpderiv{\epsilon}{\upsilon\,}{\eta}= 
    -\frac{(\gamma-1)\rho}{\beta},\qquad
    \pderivd{\epsilon}{\eta} = \frac{T}{C_\upsilon}.
\end{equation}
These equations are exact and apply to any fluid. Once an $\epsilon(\upsilon,\eta)$ is
determined the various thermodynamic parameters and the equation of state are
defined through these relations.

For an ideal gas of identical atoms/molecules with mass $m$ and $f$ degrees of freedom,
one has
\begin{equation}
    \epsilon=A \upsilon^{1-\gamma} \exp \left[ \left(\gamma-1\right)\eta/R \right],
\end{equation}
where
\begin{equation}
    \gamma = \frac{2+f}{f} \qquad
    R=\frac{k_\text{B}}{m} \qquad 
    A=\frac{3\pi\hbar^2}{e^\gamma m^{\gamma+1}}.
\end{equation}
Here $k_\text{B}$ is Boltzmann's constant, $\hbar$ is the rationalized Planck constant,
and $e=\exp(1)$. From these equations, it follows that
\begin{gather}
p\upsilon = RT \\
p\upsilon^\gamma = A\left(\gamma-1\right) \exp\left[ \left(\gamma-1\right) \eta/R
\right].
\end{gather}
The differential forms of these equations are
\begin{gather}
\frac{dp}{p} + \frac{d\upsilon}{\upsilon}= \frac{dT}{T} \\
\frac{dp}{p} + \gamma \frac{d\upsilon}{\upsilon} = \frac{\gamma-1}{R}\,d\eta.
\end{gather} 
For structureless point particles, $f=3$ and $\gamma = 5/3$. These values apply to
a plasma composed of equal numbers of protons and electrons with $m = \half (m_p + m_e)$.
For diatomic and linear polyatomic molecules, $f=5$ and $\gamma = 7/5$. These values
apply to the Earth's atmosphere where the two rotational degrees of freedom have
a thermal population but the vibrational modes are found only in their ground state. 
For a relativistic gas, or one composed entirely of
photons, $f=6$ and
$\gamma = 4/3$. 

Let $q$ (units: $\rm Joules\, kg^{-1}\, sec^{-1}$) 
be the amount of heat added to a parcel of fluid by thermal conduction. ohmic 
dissipation, viscous dissipation, or
radiative transfer. Then the three equivalent forms of the internal energy
equation are:
\begin{gather}
\Deriv{\eta}{t} = \frac{q}{T} \\[4pt]
\Deriv{p}{t} + \gamma p \Div \boldv = (\gamma-1)\rho q \\[4pt]
\Deriv{T}{t} + (\gamma-1) T\, \Div \boldv = \frac{\gamma-1}{R} q .
\end{gather}

Astrophysical fluids/plasmas often exhibit molecular formation/dissociation and
atomic ionization/recombination. These processes change the number densities of
the fluid constituents and can still be incorporated in the thermodynamic formalism
through the introduction of chemical potentials. The Saha equation is a familiar
example of how the results provided in this appendix generalize to include
these processes.

\section{Appendix: Radiation MHD}\label{app:RMHD}

After gravity, radiation typically has the next largest influence on MHD
waves in the combined photosphere/chromosphere. The radiation field may exchange
both energy and even (in hot tenuous plasmas) momentum with the material. A
steady equilibrium radiative flux $\F_0$ will also stratify a radiating
magneto-atmosphere along $\F_0$, which may be in a different direction than
the gravitational stratification along $\g$. This non-alignment leads to
equilibria which vary in all three spatial dimensions. 

Neglecting photon polarization, the specific intensity $I(\x,t;\mathbf{\Omega},\nu)$ (units:
Joules m$^{-2}$ s$^{-1}$ ster$^{-1}$ Hz$^{-1}$) satisfies the transfer equation
\begin{equation}
\frac{1}{c} \pderiv{I}{t} +  \mathbf {\Omega}\,\vdot\, \grad I = \eta
- \chi I~.
\end{equation}
Here, $\chi(\x,t; \mathbf {\Omega},\nu)$ (units: m$^{-1}$) is the opacity,
and $\eta(\x,t; \mathbf {\Omega},\nu)$ is the emissivity, $ \mathbf {\Omega}$ is a unit 
vector in three-dimensional Euclidean space, $\nu$ is the photon 
frequency, and $c$ is now the speed of light in this
appendix only \citep{Pom73aa,MihMih84aa,04castor,15hubeny,20kato}.

The coupling between the radiation field and the fluid is obtained by modifying
the fluid energy and momentum equations as follows:
\begin{equation}
    \pderiv{\rho\boldv}{t}+\Div\mathbf{T}=-\rho\grad\phi - \frac{1}{c}
    \int_0^\infty d\nu \oint d \mathbf {\Omega} ~ \mathbf {\Omega} \left( \eta - \chi
    I \right)~,
\end{equation}
\begin{equation}
    \pderiv{U}{t}+\Div\f= -
    \int_0^\infty d\nu \oint d \mathbf {\Omega} ~ \left( \eta - \chi
    I \right)~. 
\end{equation}

\subsection{RMHD Waves}
If the equilibrium radiation field and the fluid are everywhere in thermodynamic
equilibrium, then $\eta = \chi I$, and $I = B_\nu[T]$, where
\begin{equation}
I_{0}=B_\nu[T_0] = \frac{2h\nu^3}{c^2} \frac{1}{\exp(h\nu/k_\text{B}T_0)-1}~
\end{equation}
is the Planck function, $h$ is Planck's constant, and $k_\text{B}$ is Boltzmann's constant.
It follows that the temperature $T_0$ of the fluid and the radiation field is a constant
independent of position and time, and the radiative flux is everywhere zero. 

Under these circumstances Radiation Magneto\-hydro\-dynamic (RMHD) waves
${\boldv}({\x},t)$, may again be built from a weighted superposition of Fourier modes
\begin{equation}
{\V}(\k,\omega)
\exp\left[i({\k\,\vdot\, \x}-\omega\, t)\right],
\end{equation}
\begin{equation}
I_{1}(\k,\omega; \mathbf {\Omega},\nu)
\exp\left[i({\k\,\vdot\, \x}-\omega\, t)\right].
\end{equation}
The Fourier-transformed perturbed specific intensity is
\begin{equation}
I_{1}(\k,\omega; \mathbf {\Omega},\nu) = \frac{\chi}
{\chi + i \left(\k\,\vdot\,  \mathbf {\Omega}-\omega/c\right)}
\pderiv{B_\nu}{T_0} T_1~,
\end{equation}
where $T_1(\k,\omega)$ is the Fourier-transformed temperature fluctuation, and
$\chi = \chi_{0}( \mathbf {\Omega},\nu)$ is the equilibrium opacity.
The resulting dispersion relation for RMHD waves (vice MHD waves) is no longer
algebraic, but necessarily involves logarithms which result from the integration
over the solid angle d$ \mathbf {\Omega}$. Therefore, in addition to a variety of simple
poles corresponding to radiation-modified slow, Alfv\'en, and fast modes, 
there will also be a 
continuum of radiation modes associated with the branch cuts required by the 
logarithms \citep{92dzhalilov,94dzhalilov,96bogdan}.

If, on the other hand, one were to discard the transfer equation in favor of 
an Eddington-like approximation, and treat only the first- and second-angular moments of
the radiation field, then the resulting dispersion relation is again algebraic
(vice transcendental). In other words, approximating the radiation pressure by some
fraction of the radiation energy density (1/3 for the Eddington 
approximation) in turn replaces the branch-cut continuum by simple (usually complex) poles. 
Under the present circumstances, these additional simple poles are 
magnetically-modified radiation exchange, isotropization, and diffusion modes.
Strictly speaking, this approximation/replacement is accurate when the
optical depth of a wavelength is very large, or, equivalently, the wave is
optically-thick. It should be used only with extreme
care when a wave is optically-thin. The influence of the radiation field on the
MHD modes is different depending upon whether the wave is optically-thick or
optically-thin. The maximal damping always occurs in the transition region 
between optically-thick and optically-thin \citep{76kaneko,77kanekoa,77kanekob,MihMih84aa,99lowrie}.

\subsection{RMAG waves}
In stellar atmospheres there is always a sensible steady equilibrium radiative flux 
\begin{equation}
\F_0 = \int_0^\infty \!\!\!d\nu \oint \!d \mathbf {\Omega} ~\mathbf{\Omega}\, I_{0} ~.
\end{equation}
Not only does this provide an additional atmospheric stratification, 
but it also causes global
oscillations to have a mixed RMHD character---this is entirely analogous to how
gravitational stratification induces a transition between high- and low-plasma
beta layers. Below the solar photosphere, Radiation Magneto-acoustic-gravity (RMAG) waves
are optically-thick; they are optically-thin from the mid-photosphere 
through the chromosphere and corona. 

The simplest plane-parallel radiating atmosphere which supports a
gravity-aligned constant radiative flux requires the solution of Milne's integral equation
and the evaluation of the Hopf function. Therefore, progress in understanding
the nature of RMAG waves either entails a full numerical approach, or a local approximate
analysis \citep{95babaev,99spiegel,01birch}. 
For example, below the solar photosphere, local thermodynamic equilibrium
(LTE) obtains and the radiation diffusion limit is accurate; Rosseland mean
opacities may be safely employed. In the tenuous 
upper chromosphere and corona, resonance lines dominate the opacity and one often
works directly with tabulated radiative-loss functions. 

\bibliography{fred,tom}
\vspace{10pt}
This chapter contains \total{citnum}\ references.